\documentclass[10pt, a4paper]{article}
\usepackage{draft}

\newcommand{\fdud}[3]{{}_{#1}{}^{#2}{}_{#3}\,}
\newcommand{\dx}{\mathrm{d}x}
\newcommand{\ddx}{\mathrm{d}^dx}
\newcommand{\ddp}{\mathrm{d}^dp}
\newcommand{\pl}{\partial}
\newcommand{\R}{\mathbb{R}}
\newcommand{\N}{\mathbb{N}}
\newcommand{\C}{\mathbb{C}}

\newcommand{\Tr}{\mathrm{Tr}}
\newcommand{\Implies}{\qquad\Longrightarrow\qquad}
\newcommand{\Completion}{\pmb{\gamma}}
\newcommand{\Fock}{\mathfrak{F}}
\newcommand{\Base}{\mathscr{X}}
\newcommand{\WeylBundle}{\mathcal{W}}
\newcommand{\WeylAlg}{\mathcal{A}}

\begin{document}

\title{%
    A simple construction \\of Heat Kernels and conformal anomalies \\
    via Fedosov deformation quantization\\
    {\color{MidnightBlue}\adfopenflourishleft}\,
    \raisebox{1.5pt}{%
        \color{PineGreen}$\scriptscriptstyle\blacklozenge$%
    }\,
    {\color{MidnightBlue}\adfopenflourishright} 
}

\abstract{%
    We propose a new approach to the systematic computation of the heat-kernel expansion and, in particular, of conformal anomalies, based on Fedosov deformation quantization. The approach is fully covariant, applies to differential operators of arbitrary order, is amenable to efficient implementation, and yields closed-form expressions. As applications, we compute conformal anomalies in four, six, and eight dimensions, the last being a new result.
}
\author{Thomas Basile}
\author{Evgeny Skvortsov}
\emailAdd{thomas.basile@umons.ac.be}
\emailAdd{evgeny.skvortsov@umons.ac.be}
\affiliation{%
    Service de Physique de l’Univers, Champs et Gravitation,\\
    Universit\'e de Mons,\\
    20 place du Parc, 7000 Mons, Belgium
}
\maketitle

\newpage
\section*{Introduction}
\pagenumbering{arabic}
\setcounter{page}{2}

The Heat Kernel and its short-time asymptotic expansion play a central role in quantum field theory, spectral geometry, and mathematical physics. The Seeley--DeWitt--Gilkey coefficients encode local geometric information about differential operators and govern, among other things, one-loop ultraviolet divergences, effective actions, conformal and other quantum anomalies, spectral invariants, and index-theoretic quantities. Their computation has therefore been approached from many complementary directions, including the original recursive heat-kernel construction, pseudodifferential and symbol calculus, invariant-theoretic methods, normal-coordinate expansions, and worldline techniques. While these approaches are well developed, explicit calculations at high orders or for higher-order differential operators rapidly become technically involved, motivating the search for computational frameworks that retain covariance while providing a systematic and efficient expansion.

Roughly speaking, covariant Heat Kernel expansion is a difficult problem for the following reasons. At least formally, given an operator $\widehat{H}=-(\nabla^2)^\ell+\dots$ one has to compute $\Tr \exp[-t \widehat{H}]$, which involves, via the Taylor expansion, products $\widehat{H} \circ \cdots \circ \widehat{H}$ of differential operators on a manifold. The latter is difficult to find a closed canonical formula for since the covariant derivatives do not commute. As a result, one has to resort to one or another kind of expansion, e.g. by sacrificing the manifest covariance for the curvature expansion over a flat space. Another manifestation of the same problem is the lack of an efficient Fourier transform in a general curved setting,\footnote{Let us nevertheless mention the work of Widom \cite{Widom:1980mmt}, developing a covariant version of the Fourier transform in curved space, which finds its use in computing the Heat Kernel as reviewed below.} which again leads one to consider expansion over the flat space.    

The Fedosov deformation quantization \cite{Fedosov:1994zz, Fedosov:1996} solves all of the problems above by providing a simple gadget to multiply differential operators via their symbols. The symbols are `immersed' into a bigger space, the Weyl bundle over the spacetime, where $\nabla$ can be flattened out to a flat connection $D$. 
More precisely, symbols are identified with  sections of this Weyl bundle, which can be thought of as functions valued in the Weyl algebra---the algebra of symbols in flat space---and required to be covariantly constant sections with respect to $D$. One can use the Moyal--Weyl star-product inherited from flat space in a fiberwise manner to multiply these sections and by the previously mentioned identification, define a star-product for the corresponding symbol.
It is important to stress that all of this machinery does not only provide existence theorems, but concrete expressions which are covariant at all stages and that can be constructed up to the required order by simple iterations. Fedosov's original work did not give any simple formula for the trace $\Tr$, but it was constructed later by Feigin, Felder and Shoikhet \cite{Feigin:2005}. In the case of the cotangent bundle, which is the one where symbols of differential operators live, it acquires a very simple and natural form (previous work on deformation quantization of the cotangent bundle and its relation with symbol calculus include \cite{Bordemann:1997ep, Bordemann:1997er, Pflaum:1998i, Pflaum:1998ii, Pflaum:1999}, as well as the more recent \cite{Andersson:2025vxy}). The final expression looks even too simple
\begin{equation}\notag
    \int_\Base \ddx\,\sqrt{g}\,K_{\widehat f}(t,x,x)
    =\Tr\big(e^{-t\,\widehat f}\big)
    =\Tr_A\big(e_\ast^{-t\,F}\big)
    = \int_\Base\ \ddx\,\sqrt{g} \int \frac{\ddp}{(2\pi)^d}\ 
    e_\ast^{-t\,F}\Big|_{y=0}\,.
\end{equation}
Here, on the left half we have the Heat Kernel of a generic differential operator $\widehat{f}(x,\nabla)$ with symbol $f=f(x,p)$. The right half features the Fedosov lift $F=F(x,y,p)$ of $f$ with respect to Fedosov connection $A$. The connection induces the measure, which is simply $\sqrt{g}$ for cotangent bundle. Very importantly, the $p$-space, which can be thought of as an analog of the momentum space, remains flat. Since the Heat Kernel is expected to be complicated, let us point out the two main sources of complexity: the calculation of the Fedosov lift $F$ and the expansion of the star-product exponential $\exp_\ast[-tF]$ to the required order. Both can be performed efficiently by iteration. 

The main appealing feature of the approach is how easy it is to implement for a generic differential operator: there is no complicated diagrammatic or the need to find a solution to differential equations. Another nice feature is that the iterations of the Fedosov machinery are in accord with the short-time expansion of Heat Kernel. In other words, the only expansion in the paper is the small $t$-expansion itself. Since the approach gives all short-time Heat Kernel coefficients in a systematic way, it also gives conformal anomalies, which we illustrate with conformal anomalies in various dimensions. 

We attempted to make the paper mathematically sound and, hence, could not avoid terms like ``Weyl bundle'', etc. Nevertheless, we present in parallel the same story from the point of view of its practical implementation and supplement it with examples. 

The paper is organized as follows. In Section \ref{sec:HK} we review the main approaches to the Heat Kernel problem to facilitate comparison to the new method. In Section \ref{sec:Fedosov} we give a short overview of the Fedosov deformation quantization, which also introduces notation and ingredients of our approach. The main idea of the approach is discussed in Section \ref{sec:HK_expansion}, where we explain how to compute the Heat Kernel expansion for an arbitrary differential operator in a closed form and in a fully covariant manner. We present some examples in Section \ref{sec:applications}, of which the computation of the (all eleven) anomaly coefficients of the conformally-coupled scalar field in $8d$ is one genuinely new result. We discuss some future directions in Section \ref{sec:conclusions}.

\section{Heat Kernel expansion}
\label{sec:HK}

The Heat Kernel $K_{\widehat{H}}$ of a differential operator 
$\widehat{H}$ acting on functions, or even sections
of a vector bundle, on a Riemannian manifold $\Base$,
is defined as the solution of the associated `heat equation',
\begin{equation}
    \big(\partial_t + \widehat{H}_x\big)
    K_{\widehat{H}}(x,x'|t) = 0\,,
    \InEq{with}
    K_{\widehat{H}}(x,x'|t) 
        \underset{t\to0}{\longrightarrow} \delta(x,x')\,,
\end{equation}
where $\delta(x,x')$ is the Dirac distribution 
localizing at the point $x'$ on $\Base$ with respect
to the measure induced by the metric, i.e. locally
$\delta(x,x')={g(x')}^{-1/2}\delta(x-x')$.
The above heat equation can be formally solved as 
\begin{equation}
    K_{\widehat{H}}(x,x'|t) = e^{-t\,\widehat{H}_x}\delta(x,x')\,,
\end{equation}
understood as a `convolution' operator, meaning that 
it is to be integrated against functions or sections
on which $\widehat{H}$ acts over the second point $x'$.

Take for instance $\Base=\R^d$ and $\widehat{H}=-\Delta$,
the ordinary Laplacian. The formal, distributional, solution
\begin{equation}
    K_\Delta(x,x'|t) = e^{t\,\Delta_x}\,\delta(x-x')\,,
\end{equation}
can be made more explicit thanks to the existence
of the Fourier transform. Indeed, using the Fourier representation
of the Dirac distribution on $\R^d$ in the previous equation,
one finds the classical expression for the Heat Kernel
of the Laplacian,
\begin{equation}
    K_\Delta(x,x'|t) = \int_{\R^d} \tfrac{\ddp}{(2\pi)^d}\
    e^{-tp^2}e^{ip \cdot (x-x')}
    = \tfrac{1}{(4\pi t)^{d/2}}\,e^{-(x-x')^2/4t}\,,
\end{equation}
thanks to the fact that $\Delta\to-p^2$ 
under the Fourier transform, and upon performing 
a simple Gaussian integral. Unfortunately,
computing the Heat Kernel of more general operators,
on more general manifolds, is not as simple 
as this one-line computation, and one has to resort to
more sophisticated approaches. The case of the Laplace--Beltrami
operators on a compact Riemannian manifold was tackled by
Minakshisundaram and Pleijel \cite{Minakshisundaram:1949}
and later generalised by DeWitt \cite{DeWitt:1964mxt}
to Laplace-type operators (more precisely an elliptic 
second order differential operator acting on sections
of a vector bundle over a Riemannian manifold). 
The basic idea is to start from an ansatz for the Heat Kernel
near the coincidence limit, i.e. $x \to x'$, 
and for small values of $t$, that is of the same form 
as the flat space result we just derived. We will review
this in a little more details below.

Soon after, Seeley showed \cite{Seeley:1967ea}
that (modulo some additional technical assumptions
that we will not detail here for the sake of simplicity
and conciseness) in the coincidence limit,
i.e. on the diagonal $x=x'$, the Heat Kernel 
of an elliptic differential operator of order $2\ell$,
with $\ell\geq1$, admits an asymptotic, `short-time', expansion
\begin{equation}
    K_{\widehat{H}}(x,x|t)\ \underset{t\to0}{\sim}\
    \sum_{k\geq0} t^{(k-d)/2\ell}\,\mathtt{a}_k[\widehat{H}](x)\,,
\end{equation}
where the coefficients $\mathtt{a}_k$ are \emph{local}
invariants depending on $\widehat{H}$, more precisely the jet
of its symbol, often called the Seeley--DeWitt coefficients
in the physics literature. This can be turned into
an asymptotic expansion for the \emph{trace} of the Heat Kernel,
\begin{equation}\label{eq:Seeley}
    \Tr(e^{-t\,\widehat{H}}) 
    = \int_\Base \dR^dx\sqrt{g}\,K(x,x|t)
    \underset{t\to0}{\sim} \sum_{k\geq0} t^{(k-d)/2\ell}\,
    \mathtt{A}_k[\widehat{H}]\,,
\end{equation}
with
\begin{equation}\label{eq:integrated_coeffs}
    \mathtt{A}_k[\widehat{H}] := \int_\Base \dR^dx\sqrt{g}\,
    \tr\big(\mathtt{a}_k[\widehat{H}](x)\big)
\end{equation}
the integrated Seeley--DeWitt coefficients.
Let us also introduce the notation
\begin{equation}
    \Tr\big(\varepsilon\,e^{-t\widehat{H}}\big)
    \underset{t\to0}{\sim} \sum_{k\geq0} t^{(k-d)/2\ell}\,
    \mathtt{A}_k[\widehat{H}|\varepsilon]\,,
\end{equation}
for the coefficient of the trace of the Heat Kernel
with a function $\varepsilon$ inserted in.
Although the Heat Kernel is interesting on its own,
as it provides, for instance, a way of recovering
the Green's function of the operator $\widehat{H}$,
i.e. the propagator,\footnote{For example, the Heat Kernel
is used in Costello's work on homotopic renormalization
\cite{Costello:2007ei}, wherein integrating the Heat Kernel
on the `time' variable  $t$, here thought of as 
an energy/length scale, on $[\epsilon,L] \subset \R_+$
yields an `effective' propagator valid on this interval.} 
we will focus on its trace. The latter is ubiquitous in physics,
as the trace of the Heat Kernel allows one to compute
the one-loop effective action. Let us recall how this standard 
argument goes: suppose that one is interested
in a \emph{quadratic} action
\begin{equation}
    S[\Phi] = \int_\Base \dR^dx\sqrt{g}\,\Phi\widehat{H}\Phi\,,
\end{equation}
for some field $\Phi$ and a \emph{self-adjoint}
differential operator $\widehat{H}$, where for simplicity 
we are hiding any possible internal or spacetime indices 
that may be carried by $\Phi$ and the corresponding 
differential operator $\widehat{H}$,
as well as a possible trace on said implicit indices.
Thinking of this simple action as the quadratic piece
in fluctuations $\Phi$ of a non-linear action expanded
around a background, whose gravitational part is captured
by the metric $g$, one is naturally led to integrating 
the fluctuations $\Phi$ to get the one-loop effective action
\begin{equation}
    W[g,\dots] = -\ln\int \mathcal{D}\Phi e^{-S[\Phi]}
    = \tfrac12\,\ln\det(\widehat{H})\,,
\end{equation}
where the dots denote other possible background fields
than the metric. Using the Schwinger proper-time representation,
this functional can be \emph{formally} re-written as
\begin{equation}
    W[g,\dots] = -\tfrac12\int_0^\infty \frac{\dR t}{t}\,
    \Tr(e^{-t\widehat{H}})\,.
\end{equation}
This is only a formal representation of the one-loop 
effective action because the above integrand has, in general,
both UV and IR divergences, i.e. in $t\to0$ and in $t\to\infty$
respectively. The asymptotic expansion of $\Tr(e^{-t\widehat{H}})$
for $t\to0$ can then be used to extract the UV-divergent piece
of the effective action: assuming that $\widehat{H}$ 
is of second order for simplicity, we introduce a UV cut-off
$\epsilon>0$, sufficiently small so that one can use
the expansion \eqref{eq:Seeley} in the integral over $t$
on the interval $[\epsilon_0,\epsilon]$, which yields, 
\begin{equation}
    -\tfrac12\int_{\epsilon_0}^\epsilon \frac{\dR t}{t}\,
    \Tr(e^{-t\widehat{H}}) = \sum_{k=0}^{d-1} 
    \frac{\ell\,\epsilon_0^{(k-d)/2\ell}}{k-d}
    \mathtt{A}_{k}[\widehat{H}] 
    + \tfrac12\,\log(\tfrac{\epsilon_0}{\epsilon})\,
    \mathtt{A}_{d}[\widehat{H}] + (\dots)
\end{equation}
where the dots denote terms which are UV-finite, 
i.e. regular in the limit $\epsilon_0$. In practice,
one usually considers \emph{second order} elliptic operators,
of Laplace-type (meaning that are simply given by 
the Laplace--Beltrami operator plus an endomorphism, see below), 
only Heat Kernel coefficients of \emph{even order}
are non-vanishing (i.e. $\mathtt{A}_k[\widehat{H}]=0$
for $k\in2\N+1$), and as a consequence, the logarithmic piece 
appears only in even dimensions, $d \in 2\N$.
The latter is especially interesting as it turns out
to be the Weyl anomaly of the theory (if it is Weyl invariant
to begin with of course).

The Heat Kernel coefficients obey some simple properties
that one can derive from their definition as the expansion
of a trace. An important one is that,
under an infinitesimal rescaling of the operator $\widehat{H}$ 
by an arbitrary function $\varepsilon$,
the trace of its Heat Kernel varies as
\begin{equation}\label{eq:scale}
    \delta_\varepsilon \widehat{H} = \varepsilon\widehat{H}
    \Implies
    \delta_\varepsilon\,\Tr\big(e^{-t\widehat{H}}\big)
    = t\tfrac{\dR}{\dR t}\,\Tr(\varepsilon\,e^{-t\widehat{H}})
\end{equation}
from which we can readily deduce
\begin{equation}\label{eq:Weyl}
    \delta_\varepsilon \mathtt{A}_k[\widehat{H}]
    = \tfrac{k-d}{2\ell}\,\mathtt{A}_k[\widehat{H}|\varepsilon]\,.
\end{equation}
Note that all we used above is the cyclicity of the trace.

This property is particularly useful in the context of conformally
invariant theories, and conformal geometry, as it allows one 
to extract a global conformal invariant from any conformally 
covariant differential operator. Such operators are 
\begin{equation}
    \delta_\sigma g_{\mu\nu} = 2\sigma g_{\mu\nu}
    \Implies 
    \delta_\sigma \widehat{H}_g 
        = w'\sigma\,\widehat{H}_g - w\widehat{H}_g\sigma\,,
\end{equation}
where the second term should be understood as the composition
of the operator $\widehat{H}_g$ with the multiplication 
by $\sigma$, and $w, w' \in \R$ are the Weyl weights 
(typically half-integers) of the objects 
on which $\widehat{H}_g$ acts, and lands in, respectively.
Although this transformation is not exactly of the form
as \eqref{eq:scale}, it is effectively the same
\emph{under the trace}, namely the trace behaves
as if under an infinitesimal Weyl rescaling, 
the operator transforms as 
$\widehat{H}_g \leadsto \widehat{H}_g 
    + (w'-w)\,\sigma\,\widehat{H}_g$.
Consequently, one finds that the Heat Kernel coefficients 
of a conformally covariant differential operator transform
as \eqref{eq:Weyl} with $\varepsilon=(w'-w)\sigma$,
\begin{equation}\label{eq:Weyl_cov_diff}
    \delta_\sigma\mathtt{A}_k[\widehat{H}_g]
    = (w'-w)\,\tfrac{k-d}{2\ell}\,
    \mathtt{A}_k[\widehat{H}_g|\sigma]\,,
\end{equation}
and in particular, that the coefficient
of order $k=d$ is \emph{conformally invariant},
\begin{equation}
    \delta_\sigma \mathtt{A}_{d}[\widehat{H}_g] = 0\,.
\end{equation}
More generally, the fact that the Heat Kernel coefficients 
of conformally covariant operator have a fixed degree 
under \emph{infinitesimal} Weyl rescaling reflects 
the fact that they all become Weyl invariant 
in the correct dimension, and hence the integrand 
has a definite scaling behaviour modulo total derivative terms.
For instance, the coefficient $\mathtt{A}_2$
for the conformal Laplacian
$\widehat{P}_2:=\nabla^2-\tfrac{d-2}{4(d-1)}R$
is proportional to the integral of the Ricci scalar,
which transforms as
\begin{equation}
    \delta_\sigma \int_\Base \dR^dx\sqrt{g}\,R
    = (d-2)\int_\Base \dR^dx\sqrt{g}\,\sigma R
    - 2(d-1) \int_\Base \dR^dx\sqrt{g}\,\nabla^2\sigma\,,
\end{equation}
under infinitesimal Weyl transformations. The second term
is a total derivative, and hence may be discarded,
while the first one yields
\begin{equation}
    \delta_\sigma\mathtt{A}_2[\widehat{P}_2] 
    = (d-2)\mathtt{A}_2[\widehat{P}_2|\sigma]\,,
\end{equation}
in accordance with \eqref{eq:Weyl_cov_diff}
for the conformal Laplacian which is of second order,
and related scalar of Weyl weight $w=-\tfrac{d-2}{2}$
to $w'=-\tfrac{d+2}{2}$.

Let us point out that in conformal geometry, this way of deriving
conformal invariants by studying the (trace of the) Heat Kernel
of conformally covariant operators was put forward
in the late 80's / early 90's
\cite{Branson:1986, Parker:1987, Branson:1991i, Branson:1991ii, Branson:1996},
paralleling the works of physicists on conformal/Weyl anomalies
\cite{Duff:1993wm,Capper:1974ic,Deser:1976yx,Fradkin:1981jc}.

Before moving on and outlining our approach, let us briefly review
the various, standard approaches available in the literature
to compute the Heat Kernel.

\paragraph{DeWitt's approach.}
As mentioned previously, DeWitt proposed \cite{DeWitt:1964mxt}
an ansatz for the Heat Kernel of a second order
elliptic operator $\widehat{H}$ on a Riemannian manifold,

\begin{equation}\label{eq:DeWitt_ansatz}
    K_{\widehat{H}}(x,x'|t) = \tfrac{1}{(4\pi t)^{d/2}}\,
    \Delta^{\scriptscriptstyle 1/2}_{\smf VVM}(x,x')\,
    e^{-\frac{\sigma(x,x')}{2t}}\,\Xi_{\widehat{H}}(x,x'|t)\,,
\end{equation}
where $\sigma(x,x')$ is the geodesic distance between $x$ and $x'$,
often called Synge's world function, 
which used to define the Van Vleck--Morette determinant
\begin{equation}
    \Delta_{\smf VVM}(x,x') := \tfrac{1}{\sqrt{g(x)g(x')}}\,
    \det\big(\!-\tfrac{\partial}{\partial x^\mu}
    \tfrac{\partial}{\partial x'^\nu}\sigma(x,x')\big)\,,
\end{equation}
with $g(x):=\det\big(g_{\mu\nu}(x)\big)$,
and $\Xi_{\widehat{H}}(x,x'|t)$ is determined by solving 
the heat equation. Plugging the ansatz \eqref{eq:DeWitt_ansatz} 
in the heat equation, for $\widehat{H}$ of the form
\begin{equation}\label{eq:Laplace-type}
    \widehat{H} = \nabla^2 + \mathscr{E}\,,
\end{equation}
where $\mathscr{E}$ is an endomorphism of some vector bundle
$E \twoheadrightarrow \Base$ whose sections are acted 
by $\widehat{H}$,
one finds the differential equation\footnote{To do so,
one needs to use the properties $(\nabla\sigma)^2=2\sigma$
of Synge's world function, and $\Delta^{-1}_{\smf VVM}
\nabla_\mu(\Delta_{\smf VVM}\nabla^\mu\sigma)=d$
of the Van Vleck--Morette determinant.}
\begin{equation}\label{eq:heat_DW}
    \Big(\partial_t + \tfrac{1}{t}\,\nabla^\mu\sigma\,\nabla_\mu
    + \Delta_{\smf VVM}^{\scriptscriptstyle-1/2}
    \widehat{H}\,\Delta_{\smf VVM}^{\scriptscriptstyle 1/2}\Big)\,
    \Xi_{\widehat{H}} = 0\,,
    \InEq{with}
    \Xi_{\widehat{H}}(x,x'|t=0)=1\,,
\end{equation}
where all differential operators are understood 
as acting on the first argument $x$ of the biscalars.
The typical strategy here is to search for a solution
of the form 
\begin{equation}
    \Xi_{\widehat{H}}(x,x'|t) = \sum_{k\geq0} t^k\,\mathtt{b}_k[\widehat{H}](x,x')\,,
\end{equation}
which, when plugged in \eqref{eq:heat_DW}, yields the recursion
\begin{equation}
    \big(k + \nabla^\mu\sigma\nabla_\mu\big)\,\mathtt{b}_k[\widehat{H}]
    + \Delta_{\smf VVM}^{\scriptscriptstyle-1/2} 
    \widehat{H}\,\Delta^{\scriptscriptstyle 1/2}_{\smf VVM}\,
    \mathtt{b}_{k-1}[\widehat{H}] = 0\,,
    \InEq{and}
    \mathtt{b}_0[\widehat{H}] = {\cal I}\,,
\end{equation}
where ${\cal I}(x,x')$ is defined as the solution of 
\begin{equation}
    \nabla^\mu\sigma\,\nabla_\mu{\cal I} = 0
    \InEq{with}
    {\cal I}(x,x') \underset{x \to x'}{\longrightarrow} \1\,.
\end{equation}
The drawback of this method is that it heavily relies 
on the choice of ansatz \eqref{eq:DeWitt_ansatz}
which is tied to Laplace-type operators. Generalizing it
to a larger class of operators, either of higher orders
or `non-minimal' (those operators whose highest order terms
in derivative is not a power of the Laplace--Beltrami operator)
proves difficult, though some progress and a new approach 
was proposed in recent works \cite{Barvinsky:2021ijq, Barvinsky:2024kgt}. For a more detailed review,
see e.g. \cite{Barvinsky:1985, Vassilevich:2003xt, Barvinsky:2024}
or the textbooks \cite{Avramidi:2000bm, Fursaev:2011zz}.

\paragraph{Gilkey's approach.}
Gilkey's method \cite{Gilkey:1975iq, Gilkey:1980, Gilkey:2018}
for computing the Heat Kernel coefficients Laplace-type operators,
i.e. operators of the form \eqref{eq:Laplace-type},
is to use the following properties:
\begin{enumerate}[label=$(\roman*)$]
\item The coefficient $\mathtt{a}_k[\widehat{H}]$ is a linear combination
of terms of order $k/2$ in contractions of the curvature tensor
of $\nabla$ and the endomorphism $\mathscr{E}$;
\item If, on a product manifold $\Base=\Base_1 \times \Base_2$, 
the differential operator takes the form $\widehat{H} = \widehat{H}_1 + \widehat{H}_2$
where $\widehat{H}_i$ are differential operators 
acting solely on $\Base_i$, then the Heat Kernel coefficients obey
\begin{equation}
    \mathtt{a}_k[\widehat{H}]
    = \sum_{i+j=k} \mathtt{a}_i[\widehat{H}_1]\,\mathtt{a}_j[\widehat{H}_2]\,.
\end{equation}
\end{enumerate}
\vspace{-10pt}
The first property allows one to write down an ansatz
for $\mathtt{a}_k[\widehat{H}]$, enumerating all possible
contractions of the curvature and the endomorphism 
of the correct order, namely $\tfrac{k}{2}$. 
To fix the coefficients of this ansatz, one then uses 
the second property, as well as some simple identities
like \eqref{eq:Weyl}, to derive relations between them.
This method can be, and has been, extended to more general
differential operators, for instance higher order ones
\cite{Gilkey:1980, Fegan:1985}
or non-minimal ones \cite{Gilkey:1991, Avramidi:2000isc}.

\paragraph{Seeley's asymptotics.}
As mentioned previously, the asymptotic expansion 
of the Heat Kernel was derived by Seeley \cite{Seeley:1967ea},
as we now briefly review. The idea is to use the relation
between the Heat Kernel of $\widehat{H}$ to its resolvent
at the eigenvalue $\lambda$, which is nothing but the inverse
of $\widehat{H}-\lambda\1$ and from which one can recover
the heat operator $e^{-t\widehat{H}}$ via the integral transform,
\begin{equation}\label{eq:resolvent}
    e^{-t\widehat{H}} = -\int_C \tfrac{\dR\lambda}{2\pi i}\,
    e^{-t\lambda}\,\widehat{R}(\lambda)\,,
    \InEq{with}
    \widehat{R}(\lambda) := (\widehat{H}-\lambda\1)^{-1}\,,
\end{equation}
where $C$ denotes a contour encompassing the spectrum 
of the full differential operator in the complex plane.
Passing to the symbol of these operators, 
the symbol of the resolvent would be defined as the solution
of the \emph{algebraic equation},
\begin{equation}\label{eq:def_resolvent}
    \big(H(x,p)-\lambda\big) \star R(\lambda;x,p) = 1\,,
\end{equation}
where $\star$ is a star-product for the symbol map
associating the operators $\widehat{H}$ and $\widehat{R}(\lambda)$
their symbols $H(x,p)$ and $R(\lambda;x,p)$ respectively, 
which are functions on the cotangent bundle of spacetime $\Base$.
Originally, Seeley's work used only a \emph{local},
or coordinate-dependent symbol map, namely the symbol
of an operator $\widehat{H}$ is defined in terms 
of its integral kernel $\mathcal{K}_{\widehat{H}}$ via%
\footnote{Recall that the integral kernel 
$\mathcal{K}_{\widehat{H}}$ of an operator $\widehat{H}$
is defined by
\[
    (\widehat{H}\phi)(x) = \int_\Base\ddx'\,\sqrt{g(x')}
    \mathcal{K}_{\widehat{H}}(x,x')\phi(x')\,.
\]
The Heat Kernel is nothing but the integral kernel
of the heat operator $e^{-t\widehat{H}}$. 
}
\begin{equation}\label{eq:symbol}
    \mathcal{K}_{\widehat{H}}(x,x')
    = \int \tfrac{\ddp}{(2\pi)^d}\,\tfrac{1}{\sqrt{g(x')}}\,
    e^{ip \cdot (x-x')} H(x,p)\,,
\end{equation}
i.e. as its inverse Fourier transform. 
Combining this definition of symbols with that 
of the resolvent \eqref{eq:resolvent}, one ends up with
the relation
\begin{equation}
    K(x,x'|t) = -\int_C \tfrac{\dR\lambda}{2\pi i}\,
    e^{-t\lambda} \int \tfrac{\ddp}{(2\pi)^d}\,\tfrac{1}{\sqrt{g(x')}}\,e^{ip \cdot (x-x')} R(\lambda;x,p)\,.
\end{equation}
The computation of the Heat Kernel, or its asymptotics
in $t\to0$, is therefore reduced to that of the symbol
of the resolvent of $\widehat{H}$, which is the solution 
of the algebraic equation \eqref{eq:def_resolvent}.
To do so, one expands the symbol as 
$R(\lambda;x,p)=\sum_{k\geq0} R_{(k)}(\lambda;x,p)$
according to the homogeneity 
$R_{(k)}(t^{2\ell}\lambda;x,tp)=t^{-k-2\ell}R_{(k)}(\lambda;x,p)$,
and solves recursively the defining equation 
\eqref{eq:def_resolvent} order by order in this degree.

Let us stress that here, the star-product used,
the one associated with the (local) symbol map \eqref{eq:symbol},
is the normal-order star-product (also known
as the Kohn--Nirenberg star-product \cite{Kohn:1965}),
which can be derived by by computing the composing 
of the integral kernel of two operators and extracting
the resulting symbol in terms of the two operators
respective symbols. As the symbol map, it is defined 
only locally, in a chosen coordinate chart.

\paragraph{Covariant Fourier transform.}
To solve the lack of covariance problem of the previous approach,
Widom proposed a globally well-defined way of computing
the symbol of a differential operator \cite{Widom:1980mmt}
(further refined in \cite{Safarov:1997}, 
see also \cite{McKeag:2011}),
using a covariant version of the Fourier transform
based on the construction of a covariant generalization
of the flat pairing $p \cdot (x-x')$ between coordinates
and momenta. Denoting this object by $\Theta(x,x';p)$, 
one can constrain it by imposing that it be linear
in momenta, 
\begin{equation}
    \Theta(x,x';p) = p_\mu\,\Theta^{\mu}(x,x')\,,
\end{equation}
and that its symmetrized covariant derivatives verify
\begin{equation}
    \nabla_{(\mu_1} \dots \nabla_{\mu_n)}\Theta(x,x';p) 
    \underset{x \to x'}{\longrightarrow} \delta_{n,1}\,p_\mu\,,
\end{equation}
in the coincidence limit. In plain words, the first covariant
derivative of $\Theta$ at coinciding points reproduces
the momentum $p$, while itself and its n$th$ symmetrized 
derivatives for $n\geq2$ vanish at coinciding points.
This allows one to determine the repeated covariant derivatives
of $\Theta$ \emph{in the coincidence limit}: for instance,
one finds
\begin{equation}
    \nabla_\mu\nabla_\nu\Theta\rvert_{x \to x'} = 0\,,
    \qquad 
    \nabla_\mu\nabla_\nu\nabla_\rho\Theta\rvert_{x \to x'} 
    = -\tfrac{2}{3} R_{\mu(\nu}{}^\alpha{}_{\rho)}p_\alpha\,,
\end{equation}
upon rearranging the covariant derivative into totally symmetrized
pieces---which vanish in the $x \to x'$ limit---and the reminder
which necessarily involves their anticommutator, and hence
the curvature tensor. With such an object at hand, 
one defines a new symbol map via
\begin{equation}
    \mathcal{K}_{\widehat{H}}(x,x')
    = \int \frac{\ddp}{(2\pi)^d}\frac{1}{\sqrt{g(x')}}\,
    e^{i\Theta(x,x';p)}\,\widetilde{H}(x,x',p)\,,
\end{equation}
where the new symbol $\widetilde{H}(x,x';p)$ now depends
on two points in general. In the case of the resolvent,
the corresponding new symbol is constrained by the equation
\begin{equation}
    \big(e^{-i\Theta(x,x';p)}\widehat{H}e^{i\Theta(x,x';p)}
    -\lambda\big)\,\widetilde{R}(\lambda;x,x',p)
    = I(x,x')\,,
\end{equation}
with 
\begin{equation}
    \int \frac{\ddp}{(2\pi)^d}\,e^{i\Theta(x,x';p)} I(x,x')
    = \delta(x-x')\,,
\end{equation}
the covariant Fourier transform of the Dirac distribution.
One can then proceed as outlined previously, namely
expand the symbol of the resolvent in different 
homogeneous components and recursively compute them by
solving the above equation. This approach was initially 
advocated by Gusynin 
\cite{Gusynin:1989ky, Gusynin:1988zt, Gusynin:1990bu, Gusynin:1991mk} 
(with software implementation, e.g. \cite{Gusynin:1993xh}) 
and has been used in many works, see for instance the recent
\cite{Grosvenor:2021zvq}.

Having the covariantly defined symbol map \eqref{eq:symbol}
at hand, one could derive the corresponding star-product
by computing the composition of two integral kernels
in terms of said symbols, 
see e.g. \cite{Sharafutdinov:2005, Sharafutdinov:2019},
however its expression seems to become fairly complicated.
This framework is, in spirit at least, relatively close
to the one presented in the next section, in the sense
that the emphasis is put on defining and computing
the symbol of a differential operator in a `global'
or coordinate-invariant manner, and once this task is achieved,
the asymptotic expansion of the trace of the Heat Kernel
can be computed.

For additional reviews about the Heat Kernel and its application in physics, see e.g. \cite{Barvinsky:2015bky, Esposito:2015kep, Avramidi:1997jy}.

\section{Fedosov deformation quantization}
\label{sec:Fedosov}
The material below is quite standard in math, but not in physics. Therefore, let us try to explain in physics terms what is going to happen. The problem is to learn how to multiply two differential operators $\widehat{f}(x,\nabla)$, $\widehat{g}(x,\nabla)$. For example, one can choose `normal' ordering, which is to agree to keep all $x$-dependence to the left of covariant derivatives $\nabla$, e.g. $\widehat{f}(x,\nabla)=f^{\mu_1\dots\mu_k}(x)\nabla_{\mu_1 \dots \mu_k}+\dots$, where $\dots$ denotes lower derivative terms. It is important to stress that one has to choose some ordering, otherwise it is impossible to compare operators. In computing the superposition/product of two operators $\widehat{f}(x,\nabla) \widehat{g}(x,\nabla)$ the main complication comes from having to put the terms into the chosen order. In case $\nabla$ is just $\partial_\mu$, the problem is easy to solve and the solution is known as the star-product (one of its versions that corresponds to the normal ordering). First of all, if some ordering is chosen, one can `erase hats' and pass from an operator $\widehat{f}(x,\nabla)$ to its symbol $f(x,p)$, where $\nabla_\mu$ is replaced by a ordinary commuting variable $p_\mu$. Then, the star-product
\begin{equation}
    \big(f \ast g\big)(x,p) = f(x,p)\,
    \exp\Big(
    \tfrac{\overleftarrow{\partial}}{\partial p}
    \cdot \tfrac{\overrightarrow{\partial}}{\partial x}\Big)\,g(x,p)\,.
\end{equation}
produces the symbol of the product $\widehat{f}(x,\nabla) \widehat{g}(x,\nabla)$ with all terms put in the normal ordering. In practice, it is more convenient to work with the symmetric ordering and the corresponding product is called Moyal--Weyl star-product, see below.

The main problem is what to do when $[\nabla_\mu,\nabla_\nu]\neq0$. A solution by Fedosov was to `put the problem into a flat fiber'. The symbols are extended from $f(x,p)$ to $f(x;y,p)$ and the star-product is the usual Moyal--Weyl one on $y$, $p$. The original symbols of $\widehat{f}(x,\nabla)$ correspond to $f(x;0,p)$. It is also possible to uplift $\nabla$ to a flat connection $D=\nabla+\dots$ acting on such symbols, $[D_\mu,D_\nu]=0$. The dependence of symbols on $y$ is fixed from $Df=0$, i.e. by gluing different fibers via $D$. Now, the star-product $f\ast g$ of two covariantly constant sections $Df=Dg=0$ is covariantly constant again $D(f\ast g)=0$ and gives back the symbol of the wanted product of two operators. Therefore, the main trick is to flatten the connection $\nabla$ by extending the setup to a bigger space.

\paragraph{Symbols of differential operators
and quantization of the cotangent bundle.}
Fedosov approach to deformation quantization 
of symplectic manifolds \cite{Fedosov:1994zz, Fedosov:1996}
allows one to obtain \emph{globally defined} symbols 
of differential operators \cite{Fedosov:2001}.
The idea is relatively simple to summarize:
above each point, one attaches a copy of the flat model
for the algebra of symbols, thereby defining a fiber bundle
of associative algebras called the Weyl bundle. 
This bundle admits flat connections that can be built out
of not necessarily flat, ordinary, ones that can be used
to \emph{lift} functions on the cotangent bundle of $\Base$
to covariantly flat sections of the Weyl bundle,
which are globally defined. In fact, this lift defines
a bijection between covariant constant sections
of the Weyl bundle and functions on the cotangent bundle
of $\Base$. The advantage of this construction is that now,
one can use the simple and well-known tools of Weyl symbol
calculus on sections of the Weyl bundle, and in particular
define a quantization map associating functions on $T^*\Base$
to differential operators on $\Base$,
which is globally well-defined.

Let us start from the beginning, that is with defining
the Weyl bundle as
\begin{equation}
    \WeylBundle_\Base := S(T\Base) \otimes \hat S(T^*\Base)
    \twoheadrightarrow \Base\,,
\end{equation}
where $\Base$ denotes our spacetime manifold of dimension $d$,
and $\hat S(\dots)$ the completion of the symmetric algebra,
i.e. one allows elements of the tensor product to have
an infinite number of non-zero components. Concretely,
a generic section of this bundle locally reads
\begin{equation}
    \Gamma(\WeylBundle_\Base)\ \ni\ \mathscr{F}(x;y,p)
    = \sum_{k,l} \mathscr{F}_{a_1 \dots a_k}^{\,b_1 \dots b_l}(x)\,
    y^{a_1} \dots y^{a_k}\,p_{b_1} \dots p_{b_l}\,,
\end{equation}
where the lower case Latin indices $a,b=0,1,\dots,d-1$
are Lorentz indices and hence take $d$ values,
and where $\{y^a\}$ and $\{p_b\}$ define a basis
of the cotangent and tangent space over the spacetime
point $x \in \Base$, respectively. There is a slight difference
in the dependency of such sections in $y$ and $p$,
namely the latter is \emph{polynomial} while the former
is \emph{formal}, meaning sections of the Weyl bundle
are allowed to be formal power series in the $y$ variables.
This distinction reflects the different use of the symmetric
tensor algebra and its completion in the definition 
of $\WeylBundle_\Base$ above.
At each point of the base/spacetime manifold $\Base$,
the fiber over it is isomorphic, to the Weyl algebra 
$\WeylAlg_{2d}$ (modulo the subtlety that it should be extended
over $\R\llbracket\hbar\rrbracket$ so that sections
are also formal power series in $\hbar$),
which is an associative (but non-commutative) algebra,
generated by the $(d+d)$ variables $y^a$ and $p_a$,
and whose product $\ast$ is given by
\begin{equation}
    \big(\mathscr{F} \ast \mathscr{G}\big)(y,p) = \mathscr{F}(y,p)\,
    \exp\Big(\tfrac\hbar2\,
    \big[\tfrac{\overleftarrow{\partial}}{\partial y}
    \cdot \tfrac{\overrightarrow{\partial}}{\partial p}
    -\tfrac{\overleftarrow{\partial}}{\partial p}
    \cdot \tfrac{\overrightarrow{\partial}}{\partial y}\big]\Big)\,\mathscr{G}(y,p)\,.
\end{equation}
This product is called the Moyal--Weyl product,
and admits an anti-involution $(-)^\dagger$,
that is to say an operation which satisfies
\begin{equation}
    (\mathscr{F} \ast \mathscr{G})^\dagger = \mathscr{G}^\dagger \ast \mathscr{F}^\dagger\,,
    \qquad 
    (c\,\mathscr{F})^\dagger = c^*\,\mathscr{F}^\dagger\,,
    \qquad
    \forall\, \mathscr{F}, \mathscr{G} \in \WeylAlg_{2d}\,,
    \quad \forall\, c \in \C\,,
\end{equation}
where $(-)^*$ denotes the complex conjugation,
defined on generators of the Weyl algebra by
\begin{equation}
    \hbar^\dagger = -\hbar\,,
    \qquad 
    (y^a)^\dagger = y^a\,,
    \qquad 
    (p_a)^\dagger = p_a\,.
\end{equation}
In other words, it essentially boils down to complex conjugation,
upon considering the formal parameter $\hbar$
as \emph{purely imaginary}.\footnote{One could also replace
$\hbar$ with $i\,\hbar$ in the formula
for the Moyal--Weyl star-product, and consider $\hbar$
to be inert under the $\dagger$ operation. We prefer
to stick to the former convention in order to avoid 
the proliferation of imaginary factors in formulae.}
We can therefore use the Moyal--Weyl product in each fiber
to multiply sections of the Weyl bundle, so that
$\WeylBundle_\Base$ becomes a bundle of associative algebras.

Now let us construct, as announced above, a flat connection
on the Weyl bundle, starting from a spin-connection 
$\omega^{a,b} = - \omega^{b,a}$, which is torsion-free
with respect to the vielbein $e^a_\mu$, i.e. $\nabla e^a=0$,
and preserves the fiber metric $\eta^{ab}$,
i.e. $\nabla\eta^{ab}=0$. The vielbein and spin-connection
allow us to define two \emph{derivations} of the Weyl bundle,
\begin{equation}\label{eq:def_delta}
    \delta := -\tfrac1\hbar\,[{\rm d} x^\mu\,e^a_\mu\,p_a,-]_\ast\,,
    \qquad 
    \nabla := {\rm d} + \tfrac1\hbar\,
    [{\rm d}x^\mu\,\omega_\mu^{a,b}\,p_a\,y_b, -]_\ast\,,
\end{equation}
which verify
\begin{equation}
    \delta^2 = 0\,,
    \qquad 
    \delta\nabla + \nabla\delta = 0\,,
\end{equation}
so that $\delta$ is a \emph{differential}
on $\Gamma(\WeylBundle_\Base)$, that anticommutes with $\nabla$
as a consequence of the torsion-freeness of the latter.
One can easily check that
\begin{equation}
    \nabla^2 = \tfrac1\hbar\,[R^\nabla,-]_\ast\,,
    \qquad\text{with}\qquad
    R^\nabla := \big(\dR\omega^{a,b}
    + \omega^{a,}{}_c\,\omega^{c,b}\big)\,p_a\,y_b\,.
\end{equation}
We can then build a flat connection $D$ of the form
\begin{equation}\label{eq:Fedosov_connection}
    D=d+A\,, \qquad A = A_0 + \Completion\,,
    \InEq{with}
    A_0 = e^a\,p_a + \omega^{a,b}\,p_a y_b\,,
\end{equation}
and where $\Completion \in \Omega^1(\Base,\WeylBundle_\Base)$
is a $1$-form valued in the Weyl bundle, which is linear in $p$
and of order $2$ and higher in $y$, and whose components
are polynomials in the covariant derivatives of the curvature
of $\nabla$, and its contractions. More precisely, 
one can solve the flatness condition,
\begin{equation}
    \dR A + \tfrac1{2\hbar}\,[A,A]_\ast = 0
    \qquad\Longleftrightarrow\qquad 
    \delta \Completion = R^\nabla + \nabla\Completion
    + \tfrac1{2\hbar}\,[\Completion,\Completion]_\ast\,,
\end{equation}
for $\Completion$, order by order with respect to the grading
\begin{equation}\label{eq:deg}
    \deg(y)=1=\deg(\hbar)\,,
    \qquad 
    \deg(p)=0\,,
\end{equation}
which, in particular, is such that the Moyal--Weyl product
is of degree $0$. This can be done in practice by using
the contracting homotopy for $\delta$, given by
\begin{equation}\label{eq:defh&N}
    h := \tfrac1N\,y^a\,e_a^\mu\,\tfrac{\partial}{\partial (\dx^\mu)}\,,
    \qquad\qquad
    N := y^a\,\tfrac{\partial}{\partial y^a}
        + \dx^\mu\,\tfrac{\partial}{\partial (\dx^\mu)}\,,
\end{equation}
with $N$ the number operator returning the sum
of the form degree and $y$-degree of its argument. 
A simple calculation shows that it obeys
\begin{equation}
    h\,\delta + \delta\,h = \1 - i \circ p\,,
\end{equation}
where 
\begin{equation}
    i: \Functions(T^*\Base)
        \hookrightarrow \Omega(\Base,\WeylBundle_\Base)\,,
    \InEq{and}
    p: \Omega(\Base, \WeylBundle)
        \twoheadrightarrow \Functions(T^*\Base)
\end{equation}
are the canonical inclusion and projection of functions
$\Functions(T^*\Base)$ in, or from, differential forms
valued in the Weyl bundle. This allows one to find 
the completion of $A_0$ the relation
\begin{equation}\label{recurA}
    \Completion = h\big(R^\nabla + \nabla\Completion
    + \tfrac1{2\hbar}\,[\Completion, \Completion]_\ast\big)\,,
\end{equation}
upon imposing $h\Completion=0$. One can check 
that this condition can be consistently imposed,
and that it allows to compute the degree $n$
component on the left hand side in terms of 
the order $k<n$ appearing in the right hand side
of the above relation. The lowest orders reads
\begin{equation}\label{eq:connection}
    \begin{aligned}
        \Completion
        & = - \tfrac13\,\dx^\mu
        \big(R_{\mu b}{}^a{}_c\,y^b y^c 
        + \tfrac14\,\nabla_b R_{\mu c}{}^a{}_d\,y^b y^c y^d \\
        & \hspace{50pt}
        +\big[\tfrac1{20}\,\nabla_b \nabla_c R_{\mu d}{}^a{}_e
        + \tfrac{1}{15} R_{\mu b}{}^\times{}_c\,
        R_{\times d}{}^a{}_e\big]\,y^b y^c y^d y^e
        + \cdots\big)\,p_a\,.\\[3pt]
    \end{aligned}
\end{equation}

\paragraph{Practical implementation.}
In practice we note that $\Completion$ is a one-form that is linear in $p$:
\begin{align}
    \Completion&= \dx^\mu\, \gamma_\mu^m(x,y)p_m\equiv \gamma^m(x,y)p_m\,.
\end{align}
The Moyal--Weyl commutator for such symbols reduces to its Poisson bracket part and gives nothing but the Lie bracket of vector fields $\gamma^m$. Therefore, \eqref{recurA} gives
\begin{align}\notag
    \dx^\mu\, \gamma_\mu^m(x,y)p_m&= \frac{1}{N_y+1} \,y^a\,e_a^\sigma\,\tfrac{\partial}{\partial (\dx^\sigma)}\left(\nabla_\mu \gamma_\nu^m-\nabla_\nu \gamma_\mu^m +\partial_n \gamma^m_\mu\gamma^n_\nu-\partial_n \gamma^m_\nu\gamma^n_\mu\right) p_m\, \tfrac12 \dx^\mu\wedge \dx^\nu\,,
\end{align}
which we can be further simplified to
\begin{align}
    \dx^\mu\, \gamma_\mu^m(x,y)p_m
    & = \frac{1}{N_y+1}\,y^a\left(\nabla_a \gamma_\mu^m
    + \partial_n \gamma^m_a\gamma^n_\mu\right) p_m\,\dx^\mu\,,
\end{align}
upon taking into account $h\Completion=0 
\Longleftrightarrow y^a e^\mu_a\,\gamma_\mu^m(x,y) = 0$.
The recursion starts with $R^\nabla\equiv R\fdud{\mu\nu,}{a}{b} p_a y^b\tfrac12 \dx^\mu\wedge \dx^\nu$, and the first few iterations are given in \eqref{eq:connection}.

\paragraph{Extension by a vector bundle.} The same procedure as above would allow one to reconstruct
a flat connection starting from another
$1$-form connection on $\WeylBundle_\Base$ 
of the form $A_0 = \dx^\mu\,e_\mu^a\,p_a + \dots$,
where $e_\mu^a$ are the components of an invertible
vielbein on $\Base$ and the dots denote higher order terms
in $y$ and $p$. In other words, one can always find 
a $1$-form $\Completion$ valued in the Weyl algebra
so that $A=A_0+\Completion$ is flat, provided that
the piece of $A_0$ proportional to $p$ is invertible. For example, one can tensor the Weyl bundle $\WeylBundle_\Base$
with ${\rm End}(E)$ for any vector bundle
$E \twoheadrightarrow \Base$,
\begin{equation}
    \WeylBundle_\Base^E := \WeylBundle_\Base \otimes {\rm End}(E)\,,
\end{equation}
and construct a flat connection on this bundle 
starting from a torsion-free metric connection on $\Base$
and a $T\Base$-on-$E$ connection following a similar
procedure as the one outlined above. This data equips 
the Weyl bundle $\WeylBundle_\Base^E$ with connection
locally expressed as
\begin{equation}
    \nabla^E = {\rm d} + \tfrac{1}{\hbar}
    \big[\omega^{a,b} p_b y_a + \hbar\,\Gamma,-]_{\ast,E}\,,
\end{equation}
with $\Gamma$ the aforementioned connection $1$-form on $E$,
and $[-,-]_{\ast,E}$ the commutator of the associative product
in the fibres of the Weyl bundle. The latter consist
of a tensor product of the Weyl algebra in $2d$ generators
and $r \times r$ matrices, with $r={\rm rank}(E)$,
and the product is given by the Moyal--Weyl product
and extended with matrix multiplication.

The curvature of $\nabla^E$ is given by the $2$-form
\begin{equation}
    \Omega^2\big(\Base,\End(E)\big) \ni
    R^{\nabla^E} = R^\nabla\,\1_E + \hbar\,\Omega^E\,,
    \InEq{with}
    \Omega^E = {\rm d}\Gamma + \tfrac12\,[\Gamma,\Gamma]_E\,,
\end{equation}
and $R^\nabla$ as above. One can `complete' this connection
into a flat one in the Weyl bundle, whose local expression is given by
\begin{equation}
    A = A_0 + \Completion^E\,,
    \qquad 
    A_0 = e^a p_a + \omega^{a,b} p_a y_b + \hbar\,\Gamma\,,
\end{equation}
and where $\Completion^E\in\Omega^1(\Base,\WeylBundle_\Base^E)$
is a $1$-form valued in the Weyl bundle, of degree $2$ and higher,
obtained order by order via
\begin{equation}
    \Completion^E = h\big(\nabla^E\Completion^E + R^{\nabla^E}
    + \tfrac{1}{2\hbar}\,[\Completion^E,\Completion^E]_{\ast,E}\big)\,.
\end{equation}
More concretely, up to order $4$ in the degree \eqref{eq:deg},
its reads as
\begin{equation}\label{eq:connection_E}
    \begin{aligned}
        \Completion^E
        & = - \tfrac13\,\dx^\mu
        \big(R_{\mu b}{}^a{}_c\,y^b y^c 
        + \tfrac14\,\nabla_b R_{\mu c}{}^a{}_d\,y^b y^c y^d \\
        & \hspace{50pt}
        +\big[\tfrac1{20}\,\nabla_b \nabla_c R_{\mu d}{}^a{}_e
        + \tfrac{1}{15}\,R_{\mu b}{}^\times{}_c\,
        R_{\times d}{}^a{}_e\big]\,y^b y^c y^d y^e 
        + \cdots\big)\,p_a \\ 
        & \quad -\tfrac\hbar2\,\dx^\mu\big(\Omega_{\mu a}\,y^a
        + \tfrac13\,\nabla_a \Omega_{\mu b}\,y^a y^b
        + \tfrac1{24}\,R_{\mu a}{}^\times{}_b\,
        \Omega_{\times c}\,y^a y^b y^c + \dots \big)\,.
    \end{aligned}
\end{equation}
Note that, by construction, the above $2$-form is a sum
\begin{equation}
    \Completion^E = \Completion\1_E + \Completion^\Omega\,,
\end{equation}
where the term along the identity is the same $1$-form
as in \eqref{eq:connection}, defining a Fedosov connection
in the case where the endomorphism bundle $\End(E)$ is absent,
and the second term $\Completion^\Omega$ contains
both pieces of the curvature, $R_{ab}{}^c{}_d$ and $\Omega_{ab}$,
and is independent of $p$. This last property will be relevant
later on.

Lastly, let us mention that little changes in the practical implementation. One just needs to add $\hbar\,\Omega^E$ to the initial curvature $R^\nabla$ and be careful since $\Omega^E$ and its covariant derivatives, being matrix valued, do not commute to each other. For example, $\Gamma$ can be a $U(N)$ Yang--Mills gauge field with $\Omega^E$ being the corresponding field-strength.

\paragraph{Lift of symbols.}
Having equipped the Weyl bundle $\WeylBundle_\Base$,
or its extension $\WeylBundle_{\Base}^E$ by the bundle
of endomorphisms of any other vector bundle $E$,
with a flat connection, we can now lift any function
$f(x,p) \in \Functions\big(T^*\Base,\End(E)\big)$
on the cotangent bundle which is polynomial in the momenta $p_a$
and valued in endomorphisms of $E$, and thought of
as the (total) symbol of some differential operator
between sections of $E$,
to a section $F(x;y,p) \in \Gamma(\WeylBundle_\Base^E)$
in a unique manner. This is done simply by requiring $F$
to be covariantly constant with respect to the flat connection
built previously, and restricts to $f$ at $y=0$, i.e.
\begin{equation}
    \dR F + \tfrac1\hbar\,[A,F]_{\ast,E} \equiv -\delta F
    + \nabla F + \tfrac1\hbar\,[\Completion^E,F]_{\ast,E} = 0\,,
    \qquad 
    F\rvert_{y=0} = f\,.
\end{equation}
Again, using the contracting homotopy this equation can be put
in the form
\begin{equation}
    F = h\big(\nabla^E F + \tfrac1\hbar\,[\Completion^E,F]_{\ast,E}\big)\,,
\end{equation}
which allows us to solve for $F$ order by order
(with respect to the grading \eqref{eq:deg} introduced before).
This solution is unique, and therefore defines a bijection
\begin{equation}
    \begin{aligned}
        \tau: \Functions\big(T^*\Base,\End(E)\big)\
        & \overset{\sim}{\longrightarrow}\
        \Gamma_{flat}(\WeylBundle_\Base^E) \\
        f(x,p)\,\ & \longmapsto\ F(x;y,p)
                                \equiv \tau(f)(x;y,p)\,,
    \end{aligned}
\end{equation}
between flat (i.e. covariantly constant) sections
of the Weyl bundle $\WeylBundle_\Base^E$, and (total) symbols
of differential operators on $\Base$, whose first few orders read
\begin{equation}\label{eq:symbol_lift}
    \begin{aligned}
    F=\tau(f) & = f + y^a \nabla_a f
    + \tfrac12\,y^a y^b\,\big(\nabla_a \nabla_b 
    + \tfrac13\,R_{da}{}^c{}_b\,p_c
    \tfrac{\partial}{\partial p_d}\big)f
    - \tfrac\hbar4 y^a \{\Omega_{ab}, \tfrac{\partial}{\partial p_b}f\}_E\\
    & \qquad + \tfrac16\,y^a y^b y^c\,
    \big(\nabla_a \nabla_b \nabla_c
    + \big[\tfrac12\,\nabla_a R_{db}{}^e{}_c\,
    + R_{da}{}^e{}_b\,\nabla_c\big]\,p_e
    \tfrac{\partial}{\partial p_d}\big)f \\
    & \qquad - \tfrac{\hbar}{2}y^a y^b\big(\tfrac13\{\nabla_a \Omega_{bc}, \tfrac{\partial}{\partial p_c}f\}
    - \tfrac12\{\Omega_{ca}, \nabla_b\,
        \tfrac{\partial}{\partial p_c}f\}\big) 
    + \dots\,,
    \end{aligned}
\end{equation}
and whose inverse is simply obtained by evaluating the section
at $y=0$, i.e. $\tau^{-1}(F) = F\rvert_{y=0}$.%
\footnote{Note that for a simple function
$f \in \Functions(\Base)$,
i.e. a symbol that is independent on $p$, this lift
is nothing but a `covariant version' of its Taylor series,
\[
    f=f(x) \Implies \tau(f) = \sum_{n\geq0} \tfrac{1}{n!}
    y^{a_1} \dots y^{a_n} \nabla_{(a_1} \dots \nabla_{a_n)} f\,.
\]
This covariant Taylor series is simply the pullback by
the exponential map determined by the connection $\nabla$.
The lift of arbitrary symbol can be thought of as an extension
of this exponential map on $\Base$ to one on $T^*\Base$.}

\paragraph{Practical implementation.}
There is little difference to the case of $\Completion$. A more explicit formula for $F$ reads
\begin{align}
    F = \tfrac{1}{N_y} \,y^a\,e_a^\mu\big(\nabla^E_\mu F + \tfrac1\hbar\,[\Completion^E_\mu,F]_{\ast,E}\big)\,,
\end{align}
where iterations start with $f(x,p)$. It is important to remember that the commutator $[\bullet,\bullet]_\ast$ cannot be in general truncated to its Poisson bracket part already for the second order operators. Nevertheless, given an order-$2\ell$ operator, e.g. $f\sim (p^2)^\ell$, its lift remains capped at order $2\ell$.

\paragraph{Star-product.} We can then use the bijection between symbols and their lifts to define a star-product
on symbols by `pulling back' the Moyal--Weyl star-product,
\begin{equation}\label{eq:star-product}
    f \star g = \big(\tau(f) \ast \tau(g)\big)\big|_{y=0}\,,
    \qquad 
    f, g \in \Functions_{pol}\big(T^*\Base,\End(E)\big)\,,
\end{equation}
so that the associative character of $\star$ directly follows
from that of $\ast$, the associative product in the fiber
of the Weyl bundle. In a nutshell, the idea of this construction
is to use the flat model of the deformation quantization
of a symplectic manifold, that is $\R^{2d} \cong T^*\R^d$
with the Moyal--Weyl star-product, and use it in a fiberwise manner.
Indeed, the above formula tells us that we can compute
the star product of functions on the cotangent bundle
$f,g \in \Functions(T^*\Base)$ by lifting them
to flat sections of the Weyl bundle, on which we can use
the simple, flat, model for the star-product that is Moyal--Weyl's
in a fiberwise manner, and then project the result back
to the cotangent bundle by setting $y=0$.

\paragraph{Invariant trace.}
Last but not least, the Weyl bundle is equipped with a trace,
more precisely one can define a (essentially unique) trace 
on the space of its covariantly constant sections. While for a generic Fedosov connection $A$ the trace might look horrendous, see \cite{Feigin:2005,Basile:2022nou,Basile:2024hjg}, for the case of a cotangent bundle with $A$ that is linear in $p$ the final result is very simple and natural:
\begin{equation}
    \Tr_A(F) = \int_\Base \ddx\sqrt{g} 
    \int \frac{\ddp}{(2\pi)^d}\ F \Big|_{y=0}\,,
\end{equation}
where $F$ is a covariantly constant section, and the factor
$1/(2\pi)^d$ is introduced to conveniently fix the normalisation
in accordance with the standard Heat Kernel computations.

\paragraph{Fock space bundle.}
Having constructed a global lift of symbols of differential operators
on (vector bundles over) $M$, in the guise of flat sections
of the Weyl bundle, the final ingredient we need is a way
of constructing global lift of functions (or sections) 
on which the differential operators we are interested in act,
that we will use to define an action of symbols on these functions
(or sections) reproducing that of their corresponding 
differential operator. We follow closely the construction
outlined \cite[App. A]{Grigoriev:2016bzl}.

Going back to the `flat model' where symbols are elements
of the Weyl algebra, we should turn our attention 
to a space on which these symbols can act, hence a representation
of this algebra.
The natural representation of the Weyl algebra, thought of
as differential operators on $\R^d$ with polynomial coefficients,
$\WeylAlg_{2d} \cong \R[y^a,p_b]$, is the Fock representation,
thought of as polynomials (or even formal power series 
for our purpose) on $\R^d$, i.e. $\Fock_d \cong \R[y^a]$.
The action of $\WeylAlg_{2d}$ on $\Fock_d$ is given by
\begin{equation}\label{eq:Weyl_quantization}
    \big(\rho(f) \varphi\big)(y)
    = f(y,p)\,\exp\Big(\!-\hbar\,
    \tfrac{\overleftarrow{\partial}}{\partial p} \cdot 
    \big[\tfrac{\overrightarrow{\partial}}{\partial y}
    +\tfrac12 \tfrac{\overleftarrow{\partial}}{\partial y}\big]\Big)\,
    \varphi(y)\big|_{p=0}\,,
    \qquad 
    f \in \WeylAlg_{2d}\,,\ 
    \varphi \in \Fock_d\,,
\end{equation}
called the \emph{quantization map}, sending a symbol
to the corresponding differential operator on $\R^d$,
in such a way that
\begin{equation}
    \rho(f) \circ \rho(g) = \rho(f \ast g)\,,
    \qquad 
    f, g \in \WeylAlg_{2d}\,,
\end{equation}
thereby defining a representation of the Weyl algebra
as announced. Under this quantization map, $y^a$ acts 
multiplicatively on $\Fock_d$, and $p_a$ acts as 
partial derivatives,
\begin{equation}
    \rho(y^a) = y^a\,,
    \qquad
    \rho(p_a) = -\hbar\,\tfrac{\partial}{\partial y^a}\,,
\end{equation}
and polynomials in $y$ and $p$ are sent to the corresponding
polynomials in these basic operators, \emph{symmetrically ordered},
for instance,
\begin{equation}
    \rho(y^a p_b) = -\tfrac\hbar2\,
    \big(y^a\,\tfrac{\partial}{\partial y^b} + \tfrac{\partial}{\partial y^b}\,y^a\big)
    = -\hbar\,\big(y^a\tfrac{\partial}{\partial y^b}
        + \tfrac12\,\delta^a_b\big)\,.
\end{equation}

The pair $(\WeylAlg_{2d}, \Fock_d)$ together with
the Moyal--Weyl product and the quantization map reviewed here
are the basic ingredients of Weyl quantization, 
or symbol calculus. This defines a quantization of 
$\R^{2d} \cong T^*\R^d$, i.e. the cotangent bundle of flat 
$d$-dimensional space. Following the same path as before,
we will think of the Fock space $\Fock_d$ with its quantization map
as the flat model that we will use to construct a bundle from.
To be more specific, let us define the Fock bundle
\begin{equation}
    \Fock_\Base := \hat S(T^*\Base)
                    \twoheadrightarrow \Base\,,
\end{equation}
whose sections are
\begin{equation}
    \Gamma(\Fock_\Base)\, \ni\, \Phi(x;y)
    = \sum_{k\geq0} \tfrac1{k!}\,\Phi_{a_1 \dots a_k}(x)\,
    y^{a_1} \dots y^{a_k}\,,
\end{equation}
that we shall extend as formal power series in $\hbar$.
In plain words, these sections are functions on $\Base$
valued in the Fock space $\Fock_d$, and as such can be acted upon
by the covariant derivative associated with the Fedosov connection 
\eqref{eq:Fedosov_connection} constructed previously, 
using the quantization map, i.e.
\begin{equation}\label{eq:cov_Fock}
    \mathfrak{D} = \dR + \tfrac1\hbar\,\rho(A)\,.
\end{equation}
In particular,
\begin{equation}
    \dR + \tfrac{1}{\hbar}\rho(A_0) 
        = -e^a\,\tfrac{\partial}{\partial y^a}
        + \big(\dR - \omega^{a,}{}_b\,
        y^b \tfrac{\partial}{\partial y^a}\big)
    \equiv -\delta + \nabla\,,
    \qquad\text{with}\qquad 
    \delta = e^a\,\tfrac{\partial}{\partial y^a}\,,
\end{equation}
so that the covariant derivative starts with the nilpotent
piece $\delta$ of degree $-1$ as previously, followed by
the ordinary covariant derivative $\nabla$ 
which is of degree $0$. More generally, the quantization map
is of degree $0$ with respect to the grading \eqref{eq:deg}
and hence the components of higher degree in the Fedosov
connection, i.e. those contained in $\Completion$,
contribute to pieces of strictly positive degree
in the covariant derivative \eqref{eq:cov_Fock}.

We can therefore solve for flat sections of the Fock bundle
in a similar manner as for the Weyl bundle, namely
\begin{equation}
    \mathfrak{D}\Phi = 0\,,
    \qquad\text{with}\qquad 
    \Phi\big|_{y=0} = \phi\,,
\end{equation}
can be solved via
\begin{equation}\label{eq:rec_phi}
    \Phi = h\Big(\nabla\Phi
        + \tfrac1\hbar\,\rho(\Completion)\Phi\Big)\,.
\end{equation}
and whose evaluation order by order again yields 
a recursive definition of $\Phi$.
As before, the covariantly constant section thus constructed 
only depends on $\phi$, sitting  on $\Base$, and hence 
we obtain a bijection
\begin{equation}
    \begin{aligned}
        \tau: \Functions(\Base)\
        & \overset{\sim}{\longrightarrow}\
        {\rm Ker}(\mathfrak{D}) \subset \Gamma(\Fock_\Base) \\
        \phi(x)\,\ & \longmapsto\ \Phi(x;y)
                                \equiv \tau(\phi)(x;y)\,,
    \end{aligned}
\end{equation}
between functions $\Functions(\Base)$ on our base manifold,
i.e. spacetime, and flat sections of the Fock bundle,
which are of the form
\begin{equation}\label{eq:lift_phi}
    \Phi(x;y) = \phi + y^a\nabla_a\phi
    + \tfrac12\,y^a y^b\,\big(\nabla_a\nabla_b
    - \tfrac16\,R_{ab}\big)\phi + \dots
\end{equation}
where $R_{ab}$ denotes the Ricci tensor of $\nabla$,
and the dots denote terms of order $3$ or higher in $y$.

With all of this in place, we now have a way to associate
to a symbol $f(x,p)$ an operator $\widehat{f}$, namely 
by defining
\begin{equation}
    \big(\widehat f\phi\big)(x)
        := \rho\big[\tau(f)\big] \tau(\phi) \rvert_{y=0}
        \equiv \rho(F)\Phi\rvert_{y=0}\,,
\end{equation}
i.e. the action of the operator $\widehat{f}$ on a function $\phi$
is computed by first lifting both the symbol $f$ and $\phi$
to a flat section of the Weyl bundle $F$, and of the Fock bundle
$\Phi$, respectively, using the Weyl quantization map
\eqref{eq:Weyl_quantization} to act with $F$ on $\Phi$,
thereby producing another flat section of the Fock bundle,
which is projected on the base by setting $y=0$.
As one would expect, momenta $p_a$ are quantized
to the covariant derivative, 
\begin{equation}
    \widehat{p}_a = -\hbar\,\nabla_a\,,
\end{equation}
and polynomials of higher order in momenta acquire
curvature corrections, thereby reflecting the non-commutative
nature of $\nabla$. For instance, if $\nabla$ is the Levi--Civita
connection associated with a metric $g_{ab}$,
then $p^2 = g^{ab} p_a p_b$ is quantized to
\begin{equation}\label{eq:p^2}
    \widehat{(p^2)} = \hbar^2\big(\nabla^2 - \tfrac{R}{4}\big)\,.
\end{equation}
This illustrates the fact that, as stressed above, 
the composition of differential operator is captured 
by the \emph{star-product} of their symbol, not the ordinary
pointwise product:
\begin{equation}
    \hbar^2 \nabla^2 = g^{ab}\,
    \widehat{p}_a \circ \widehat{p}_b
    = g^{ab} \widehat{p_a \star p_b} \neq
    \widehat{(p^2)}\,.
\end{equation}

Paralleling the case of the Weyl bundle,
one can add `coefficients' to the Fock bundle by tensoring 
it with another vector bundle $E \twoheadrightarrow \Base$,
and lift its sections to flat sections
\begin{equation}
    \Fock_\Base^E := \Fock_\Base \otimes E\,,
\end{equation}
with respect to the flat connection of $\WeylBundle_\Base^E$
constructed in \eqref{eq:connection_E}. In doing so,
one extends the quantization map by letting the matrix factors
on Weyl algebra elements act on elements of $E$, i.e.
\begin{equation}
    \rho_E(f \otimes \mathfrak{E})(\varphi \otimes e)
    := \rho(f)\varphi \otimes \mathfrak{E} e\,.
\end{equation}
Due to the split of the $\Completion^E$ into the piece 
$\Completion$ remaining even in the absence 
of the vector bundle $E$, and another piece $\Completion^\Omega$,
the lift of section of $E$ defined by constructing a flat section
of $\Fock_\Base^E$ can be simplified: indeed, one finds
\begin{equation}
    h\big(\rho(\Completion^\Omega)\Phi\big)
    = h(\Completion^\Omega)\Phi= 0\,,
\end{equation}
as a consequence of the fact that $\tfrac{\partial}{\partial p_a}\Completion^\Omega=0$,
and hence the lift $\Phi$ is obtained by the recursion
$\Phi=h\big(\nabla\Phi+\rho(\Completion)\Phi\big)$.
In other words, the lift is given by same formula 
as if the bundle $E$ was absent, 
upon replacing $\nabla\to\nabla^E$.

\paragraph{Practical implementation.}
Suppose we are interested in an operator $\widehat{f}(x,\nabla)$. 
We need its symbol $f(x,p)$ and the corresponding Fedosov lift $F(x,y,p)$: 
\begin{align}
    &\text{operator } \widehat f(x,\nabla) && \Longleftrightarrow && \text{symbol } f(x,p) &&\Longleftrightarrow && \text{Fedosov lift } F(x,y,p)\,.
\end{align}
The consistency condition is simply that the action of the lift $F$ of symbol $f$ via the quantization map gives the action of $\widehat{f}$:
\begin{equation}
    \big(\widehat{f}\phi\big)(x)
        := \rho(F) \Phi \big|_{y=0}\,,
\end{equation}
where $\phi$ is a field on which $\widehat{H}$ acts and $\Phi$ is its Fedosov lift. Given an operator $\widehat{f}$, roughly speaking, $\nabla$ goes over into $p$ up to $[\nabla,\nabla]\sim R$ terms. Therefore, it is easy to write down the most general ansatz for $f$. Next, one lifts $f$ to $F$ to the order such that higher orders do no affect $\rho(F) \Phi \big|_{y=0}$. Lastly, one evaluates $\rho(F) \Phi \big|_{y=0}$ and fixes the free coefficients. For example, suppose we are interested in $\widehat{f}$ being the conformal Laplacian:  
\begin{equation}
    \widehat{f}= \hbar^2\big(\nabla^2 
    - \tfrac{d-2}{4(d-1)}\,R\big)\,.
\end{equation}
The most general ansatz for the symbol is very short
\begin{equation}
    f = p^2 + \alpha R\,,
\end{equation}
where $\alpha$ is a constant to be fixed.
The relevant pieces of the lifts are
\begin{align}
    F & = p_ap_b\eta^{ab} + \tfrac13\,y^c y^d\,
    R_c{}^a{}_d{}^bp_ap_b + \alpha \,R + \dots\,,\\
    \Phi &= \phi + y^a\nabla_a\phi
    + \tfrac12\,y^a y^b\,\big(\nabla_a\nabla_b
    - \tfrac16\,R_{ab}\big)\phi + \dots\,,
\end{align}
and the relevant evaluation of the quantization map are
\begin{equation}\label{eq:q_lem}
    \rho(p^2)\rvert_{y=0} = \hbar^2\,\partial_y^2\,,
    \qquad\qquad
    \rho(y^a y^b p_c p_d)\rvert_{y=0}
    = \tfrac{\hbar^2}{2}\,\delta^{(a}_c\,\delta^{b)}_d\,, \qquad\qquad \rho(1)=1\,.
\end{equation}
Therefore, the action of $F$ reads
\begin{equation}
    \widehat{f}\phi = \hbar^2\,\big(\nabla^2
    +[\tfrac{\alpha}{\hbar^2} -\tfrac14]\,R\big)\phi\,,
\end{equation}
which implies $\alpha = \tfrac{\hbar^2}{4(d-1)}$ and, hence, the symbol of the conformal Laplacian is
\begin{equation}
    f = p^2 + \tfrac{\hbar^2}{4(d-1)}\,R\,,
\end{equation}
which we will use in what follows. This is also in accordance with the previous remark around \eqref{eq:p^2} that the quantization of $p^2$ is $\nabla^2-\tfrac{R}{4}$, and that $p$-independent symbols act multiplicatively. For operators with some special properties, e.g. conformal invariance like in the example above, there can be other, ``intrinsic'' ways to fix the symbol without having to evaluate its action, see e.g. \cite{Basile:2024hjg}.

\section{Heat Kernel expansion}
\label{sec:HK_expansion}
We can now turn our attention to the main goal: the trace of the Heat Kernel for a differential operator $\widehat{f}(x,\nabla)$ with symbol $f=f(x,p)$ whose Fedosov lift is $F=F(x,y,p)$. The measure computed for the Fedosov connection linear in $p$ just gives $\sqrt{g}$ and the final result is
\begin{equation}\label{bestHeatKernel}
    \int_\Base \dR^dx\,\sqrt{g}\,K_{\widehat f}(t,x,x)
    =\Tr\big(e^{-t\,\widehat f}\big)
    =\Tr_A\big(e_\ast^{-t\,F}\big)
    = \int_\Base\ \ddx\,\sqrt{g}
    \int \frac{\ddp}{(2\pi)^d}\, e_\ast^{-t\,F}\Big|_{y=0}\,.
\end{equation}
In principle, this is a closed formula amenable to systematic short time expansion, as we will demonstrate, since it looks deceptively tautological. There are two sources of complexity. Firstly, one has to compute to a high enough degree the Fedosov lift $F$ of the symbol $f$ corresponding to operator $\widehat{f}$. Secondly, the star-exponent needs to be expanded in powers of $\hbar$, which we will discuss shortly. The rest are purely elementary manipulations. 

In order to extract the dependency
of the integrand on $t$, let us assume that our symbol
is homogeneous, i.e. of the form
\begin{equation}\label{Fexpansion}
    F(y,p;\hbar) = \sum_{k=0}^{2\ell} \hbar^{k}\,
    F_k^{a_1 \dots a_{2\ell-k}}(y)\,
    p_{a_1} \dots p_{a_{2\ell-k}}
    \InEq{i.e.}
    F(y,tp;t\hbar) = t^{2\ell}\,F(y,p;\hbar)\,.
\end{equation}
For many interesting cases, e.g. GJMS operators, the symbol is even in $p$ and $\hbar^2$, which is signified in the index $2k$. However, odd powers of $\hbar$ are present when there is a vector bundle component, see \eqref{eq:symbol_lift}. In those cases one can assume that the sum is over $k \in \mathbb{N}/2$. We have also assumed $F$ starts with a term of order $2\ell$ in $p$. 

The next step is to tackle $\exp_\ast[\dots]$. We will see in Section \ref{sec:star-exp} that the star-product exponent can be expanded in $\hbar$ to give
\begin{equation}
    e_\ast^{\tau F} = e^{\tau F_0}\,{\cal E}(\tau)\,,
    \InEq{with}
    {\cal E}(\tau) = \sum_{k\geq0} \hbar^k\,{\cal E}_k(\tau)\,, \qquad {\cal E}_0=1\,,
\end{equation}
where ${\cal E}_k(\tau)$ are functions of $F_i$, which are polynomials in $\tau$. Upon making the change of variables
$p \to p' = t^{1/2\ell}\,p$, the above trace reads
\begin{equation}
    \Tr_A\big(e_\ast^{-t\,F}\big)
    = \int_\Base\ \sum_{k\geq0} t^{(k-d)/2\ell}\,
    \int \frac{\ddp}{(2\pi)^d}\
    e^{-F_0}\,{\cal E}_{k}(-1)\Big|_{y=0}\,,
\end{equation}
upon using the fact that the symbol $F$ is homogeneous
in the sense of \eqref{Fexpansion}.
The crucial simplification is that the $p$-space is flat and all $\ddp$ integrals are exactly doable!

Let us first focus on Laplacian-type operators, 
i.e. symbol of the form $f=p^2 + \hbar^2 \varphi(x)$.
According to the previous result, the star-exponential
of the lift of such symbols will be of the form
\begin{equation}
    e_\ast^{-t F}\rvert_{y=0} = e^{-t p^2}\,
    {\cal E}(-t)\rvert_{y=0}\,,
\end{equation}
with ${\cal E}$ a formal power series in $\hbar$
with coefficients being polynomial in $p_a$. This means that
its integral over $p_a$ is Gaussian, so that
\begin{align}\label{eq:trace_Laplace-type}
    \Tr_A\big(e_\ast^{-tF}\big) 
    & = \,\int_\Base \ddx\,\sqrt{g}\,
    \sum_{k=0}^\infty t^{(2k-d)/2}\,\int \frac{\ddp}{(2\pi)^d}\
    e^{-p^2}\,{\cal E}_{2k}(-1)\rvert_{y=0}\,,
\end{align}
where we have also used the fact that ${\cal E}_{2k+1}=0$
for all $k\geq0$ in the case of Laplace-type operators,
as will be justified in the next section.
This can be compared with the usual short-time expansion
of the trace of the Heat Kernel for Laplacian-type operators
\cite{Gilkey:1975iq, Gilkey:2018}, recalled in \eqref{eq:Seeley},
and leads us to the expression
\begin{align}
    \mathtt{a}_{2k}[\widehat{f}\,] 
    & = \int \frac{\ddp}{(2\pi)^d}\ 
    e^{-p^2}\,{\cal E}_{2k}(-1)\rvert_{y=0}\,,
    &&& \mathtt{A}_{2k}[\widehat{f}\,] & = 
    \int_\Base \ddx\,\sqrt{g} \int \frac{\ddp}{(2\pi)^d}\ 
    e^{-p^2}\,{\cal E}_{2k}(-1)\rvert_{y=0}\,,
\end{align}
for the un/integrated Seeley--DeWitt coefficients.

\subsection{Star-exponential}
\label{sec:star-exp}
The goal of this section is to compute the star-exponent of an arbitrary symbol $F(x,y,p)$. From the practical implementation point of view, the discussion below gives the most optimal way to compute it as well. Let us define the star-exponential as the power series,\footnote{The beginning of the discussion is very close to \cite{Segal:2002gd}.} 
\begin{equation}
    e_\ast^{\tau F} := \sum_{k\geq0} \tfrac{\tau^k}{k!}\,
    \underbrace{F \ast \dots \ast F}_{k\,\text{times}}\,,
\end{equation}
It will be confirmed in a bit that the $e_\ast^{\tau F}$ can be expanded in powers of $\hbar$ as follows
\begin{equation}
    e_\ast^{\tau F} = e^{\tau F_0}\,{\cal E}(\tau)\,,
    \InEq{with}
    {\cal E}(\tau) = \sum_{k\geq0} \hbar^k\,{\cal E}_k(\tau)\,, \qquad {\cal E}_0=1
\end{equation}
The defining equations for $e_\ast^{\tau F}$ is
\begin{equation}\label{eq:ODE_exp}
    \tfrac{\dR}{\dR \tau} e_\ast^{\tau F}
    = F \ast e_\ast^{\tau F}\,,
\end{equation}
It can be translated into a series of equations on ${\cal E}_k$.
Recall that the Moyal--Weyl star-product is a formal power series
in $\hbar$ with coefficients in bidifferential operator,
that take the form
\begin{equation}
    \ast = \sum_{k\geq0} \hbar^k\,m_k\,,
    \qquad\text{with}\qquad
    m_k(f,g) := \tfrac1{2^k}\,
    \sum_{j=0}^k \tfrac{(-1)^j}{j!(k-j)!}\,
    \partial_{a_1 \dots a_{k-j}} \partial^{b_1 \dots b_j} f\,
    \partial^{a_1 \dots a_{k-j}} \partial_{b_1 \dots b_j} g\,,
\end{equation}
where $\partial_a = \tfrac{\partial}{\partial y^a}$
and $\partial^a = \tfrac{\partial}{\partial p_a}$\,.
Its antisymmetry implies that 
\begin{equation}
    m_k(f,g) = (-1)^k\,m_k(g,f)\,,
\end{equation}
so that only odd/even power of $\hbar$ participate
to the star-anti/commutator,
\begin{equation}
    \tfrac12\,[-,-]_\ast \equiv \sum_{k\geq0} \hbar^{2k+1}\,m_{2k+1}(-,-)\,,
    \qquad 
    \tfrac12\,\{-,-\}_\ast \equiv \sum_{k\geq0} \hbar^{2k}\,m_{2k}(-,-)\,.
\end{equation}
Since $F \ast e_\ast^{\tau F} = e_\ast^{\tau F} \ast F$,
the defining differential equation reads
\begin{equation}
    \tfrac{\rm d}{{\rm d}\tau} e_\ast^{\tau F}
    = \tfrac12\,\big\{F, e_\ast^{\tau F}\big\}_\ast
    \equiv F\,e_\ast^{\tau F} + \mathfrak{m}(F,e^{\tau F}_\ast)
    \InEq{where}
    \mathfrak{m}(-,-) := \sum_{k\geq1} \hbar^{2k}\,m_{2k}(-,-)\,,
\end{equation}
and, with initial condition
$e_\ast^{\tau F}\underset{\tau\to0}{\longrightarrow}\ 1$,
is solved by
\begin{equation}
    e^{\tau F}_\ast = e^{\tau F}\,
    \bigg[1+\int_0^\tau \dR u\,e^{-u F}\,
    \mathfrak{m}\big(F,e^{u F}_\ast\big)\bigg]\,.
\end{equation}
Assuming that $F=\sum_{k\geq0} \hbar^k\,F_k$
is also a formal power series in $\hbar$, this equation 
can be used to extract $e^{-\tau F}\,e^{\tau F}_\ast$
order by order in $\hbar$, recursively. However, the fact
that the exponential contains a factor that is inhomogeneous
in $\hbar$ makes the expansion of the above equation in $\hbar$
a little bit messy, so we find it more convenient to only extract 
the ordinary exponential of $F_0$ from the star-exponential,
and write it as
\begin{equation}
    e_\ast^{\tau F} = e^{\tau F_0}\,{\cal E}(\tau)\,,
    \InEq{with}
    {\cal E}(\tau) = \sum_{k\geq0} \hbar^k\,{\cal E}_k(\tau)\,,
\end{equation}
so that, 

\begin{equation}\label{eq:rec}
    {\cal E}_k(\tau) = \delta_{k,0}
    + \sum_{\substack{2l+l' \leq k \\ (l,l') \neq (0,0)}}
    \int_0^\tau \dR u\ e^{-u F_0}\,m_{2l}\big(F_{l'}, 
    e^{u F_0}{\cal E}_{k-l'-2l}(u)\big)\,.
\end{equation}
Schematically, the above integral is of the form
\begin{equation}
    \int_0^\tau \dR u\, e^{-u A} B e^{u A} g(u)\,,
\end{equation}
for two operators $A$ and $B$ acting on $g$,
which is a polynomial in the integration variable.
Here $A$ is the multiplication by $F_0$,
and $B$ is the differential operator acting
on the second argument of $m_{2l}$. Since $g$ is a polynomial,
we may focus on the operator-integral appearing 
for an order $m$ monomial,
\begin{align}
    \int_0^\tau \dR u\, u^m e^{-u A} B e^{u A}
    & = \int_0^\tau \dR u\,u^m\,e^{-u[A,-]} B 
    = \tau^{m+1} \sum_{k\geq0} 
        \tfrac{(-\tau\,[A,-])^k}{k!(k+m+1)}\,B\\ \nonumber
    & = \tfrac1{N_\tau} e^{-\tau[A,-]}\,\tau^{m+1}B
    \equiv \tfrac1{N_\tau}\tau^{m+1}\,e^{-\tau A}Be^{\tau A}\,,
\end{align}
where $N_{\tau}:=\tau\partial_\tau$ checks the homogeneity 
in $\tau$. Applying this formula to the recursion \eqref{eq:rec},
we can re-write it as
\begin{align}\label{eq:recursion_exp}
    {\cal E}_k(\tau) & = \delta_{0,k}
    + \tfrac{1}{N_\tau}\,\tau
    \sum_{\substack{2l+l' \leq k \\ (l,l') \neq (0,0)}}
    e^{-\tau F_0} m_{2l}\big(F_{l'}, 
        e^{\tau F_0}{\cal E}_{k-l'-2l}(\tau)\big)\,.
\end{align}
In this form, it becomes relatively easy to identify 
the contribution to each power of $\tau$. The first few orders in $\hbar$ 
are given by
\begin{align}
    {\cal E}_1(\tau) & = \tau F_1\,, \\
    {\cal E}_2(\tau) & = \tau F_2 + \tfrac{\tau^2}{2} F_1^2
    + \tfrac{\tau^2}{8}\big[\partial_{ab} F_0 \partial^{ab} F_0
    - \partial_a \partial^b F_0 \partial^a \partial_b F_0\big] \\
    \nonumber & \hspace{50pt} 
    + \tfrac{\tau^3}{24}\big[\partial_{ab} F_0
    \partial^a F_0 \partial^b F_0 
    - 2\,\partial_a \partial^b F_0\partial^a F_0 \partial_b F_0
    + \partial^{ab} F_0 \partial_a F_0 \partial_b F_0\big]\,,\\
    \nonumber
    {\cal E}_3(\tau) & = \tau F_3 + \tau^2\Big(F_1 F_2
    + \tfrac18\big[\partial_{ab}F_0 \partial^{ab}F_1 
    - 2\partial_a\partial^bF_0\partial^a\partial_bF_1
    +\partial^{ab}F_0\partial_{ab}F_1\big]\Big)\\
    \nonumber & \quad + \tau^3\Big(\tfrac16 F_1^3
    + \tfrac18 F_1\big[\partial_{ab}F_0 \partial^{ab}F_0 
    - \partial_a\partial^bF_0\partial^a\partial_bF_0\big] \\
    & \qquad\quad + \tfrac1{24}
    \big[\partial_{ab}F_1 \partial^a F_0 \partial^b F_0
    - 2\partial_a\partial^b F_1 \partial^a F_0 \partial_b F_0
    +\partial^{ab} F_1 \partial_a F_0 \partial_b F_0 \\
    \nonumber & \qquad\quad 
    + 2\partial_{ab} F_0 \partial^a F_0 \partial^b F_1
    -2\partial_a\partial^bF_0(\partial_b F_0 \partial^a F_1
    +\partial^a F_0 \partial_b F_1)
    +2 \partial^{ab}F_0\partial_aF_0\partial_bF_1\big]\Big)\\
    \nonumber & \quad + \tau^4\Big(\tfrac1{24} F_1
    \big[\partial_{ab}F_0 \partial^aF_0 \partial^b F_0 
    - 2\partial_a\partial^b F_0 \partial^a F_0 \partial_b F_0
    + \partial^{ab} F_0 \partial_a F_0 \partial_b F_0\big]\Big)\,.
\end{align}
Note that for Laplace-type operators, $F=F_0+\hbar^2 F_2$,
and because $F_1=0$, and one can check that ${\cal E}_{2k+1}=0$
for all $k\geq0$ (this follows simply for the recursive
definition of ${\cal E}_k$ given above).

\subsection{Wick contractions}
\label{sec:Wick}
At this point the Heat Kernel expansion is reduced to  formula \eqref{eq:trace_Laplace-type}, which we reproduce here,
\begin{equation}
    \Tr_A\big(e_\ast^{-t\,F}\big)
    = \int_\Base\ddx\sqrt{g} \sum_{k\geq0} t^{(k-d)/2\ell}\,
    \int \frac{\ddp}{(2\pi)^d}\ e^{-F_0}\,{\cal E}_{k}(-1)\rvert_{y=0}\,,
\end{equation}
where $F_0\sim (p^2)^\ell$. It is easy to see that the functions ${\cal E}_{k}(-1)\rvert_{y=0}$ are polynomials in $p$ as long as $F$ was a polynomial in $p$ at each power of $\hbar$, which is what we assumed via \eqref{Fexpansion}. This is true for polynomial symbols $f(x,p)$, i.e. for operators of finite order. In other words, we have
\begin{align}
    \mathcal{E}_k(-1)&= \sum^{\text{finite}} \mathcal{E}_k^{b_1 \dots b_n} p_{b_1} \dots p_{b_n}\,.
\end{align}
For $\ell=1$, i.e. $F_0=g^{ab} p_a p_b$ one can use the Wick's theorem either in the form
\begin{equation}
    \int \ddp\,e^{-p^2}(\dots)
    = \pi^{d/2}\,e^{\frac14\,\Box_p}(\dots)\rvert_{p=0}\,,
\end{equation}
or, more generally,
\begin{align}    
\int {\rm d}^dp\,e^{-\frac{t}2 g^{ab} p_a p_b}(\dots)
    = \sqrt{\frac{(2\pi/t)^d}{|\det(g)|}}\,
    \exp\Big(\tfrac1{2t}\,g_{ab}\,
    \tfrac{\partial^2}{\partial p_a \partial p_b}\Big)
    (\dots)\rvert_{p=0}\,.
\end{align}
One of the crucial simplifications is that the ``momentum'' space, the space of $p$, is still a flat space even when the base manifold is curved!

For higher order differential operators the complexity is due to the appearance of non-Gaussian integrals with measure $\exp[-t F_0]$, $F_0=F^{a_1...a_k} p_{a_1}...p_{a_k}$.\footnote{Non-Gaussian integrals can sometimes be computed in a closed form and have a beautiful mathematics behind, see e.g. \cite{Morozov:2009kc} for a review. However, it is not our fight and the good news is that the ``momentum'' space is still flat.} Nevertheless, with the most interesting case $F_0\sim (p^2)^\ell$ the generalization is straightforward:
\begin{align}\label{higherorder}
    \int {\rm d}^dp\,e^{-(p^2)^\ell} 
    & = \frac{\pi^{d/2}}{\ell\,\Gamma(d/2)}
    \Gamma\!\left(\frac{d}{2\ell}\right)\,.
\end{align}
It is clear at this point that the dependence of the Heat Kernel coefficients on dimension $d$ of the spacetime is very simple for the usual Gaussian case: there is only an overall numerical $\pi^{d/2}$-factor from the integral and an overall $(2\pi)^{-d}$ from the normalization. Note that $F$ itself can carry an explicit $d$-dependence. For the non-Gaussian case, i.e. higher order operators, an analog of Wick's theorem is easy to develop, but the sum over pairwise contractions via the inverse metric is accompanied by a factor that depends both on $d$ and on the number of $p$-factors being contracted in a nontrivial way. Therefore, there is no $d$-independence in the higher order case. As is clear from the procedure, until we get to compute Wick contractions there is no $d$-dependence in the approach, which can be attributed to the fact that we only need the canonical symplectic geometry of cotangent bundle rather than the metric structure of the underlying spacetime manifold.

\section{Applications}
\label{sec:applications}
In this section we consider some simple and not so simple examples illustrating our approach. We mostly concentrate on the case of conformal anomalies of the conformally coupled scalar field. Given the $d$-universality present in this case one could discuss all Heat Kernel coefficients together, but it makes sense to present $\mathtt{a}_{2k}$ as the conformal anomaly in $d=2k$. We distinguish between the integrated Heat Kernel coefficients $\mathtt{A}_k$ and the corresponding densities $\mathtt{a}_k$ as in \eqref{eq:integrated_coeffs}.

\subsection{First few terms}
\label{sec:first_terms}
The first few terms are easy to get for a large class of operators. In the case of Laplacian-type operator, i.e. operators
of the form 
\begin{equation}
    \widehat{H} = \nabla^2 + \mathscr{E}\,,
    \InEq{with}
    \mathscr{E} \in \Gamma\big({\rm End}(E)\big)\,,
\end{equation}
their symbols are given by
\begin{equation}
    f = p^2\,\1_E + \hbar^2(\mathscr{E}+\tfrac{R}4\,\1_E)\,,
\end{equation}
whose lift read
\begin{equation}
    \begin{aligned}
        F & = \big(\eta^{ab}
        + \tfrac13\,y^c y^d\,R_c{}^a{}_d{}^b
        + \tfrac1{20}\,y^c y^d y^e y^f\,
        \big[\nabla_c \nabla_d R_e{}^a{}_f{}^b
        + \tfrac{4}{3}\,R^a{}_{cd\times} 
            R^b{}_{ef}{}^\times\big] + \dots\big) p_a p_b\\
        & \quad - \hbar\,\big(\Omega_{ab} y^a p^b
        + \tfrac23\,\nabla_a \Omega_{bc} y^a y^b p^c + \dots\big) \\
        & + \hbar^2\,\big(\mathscr{E}+\tfrac{R}4
        + y^a \nabla_a [\mathscr{E}+\tfrac{R}4]
        + \tfrac12\,y^a y^b\,[\nabla_a \nabla_b\mathscr{E}
        +\tfrac14\nabla_a \nabla_bR
        + \tfrac12\,R_a{}^{cde} R_{bcde}] + \dots\big)\,.
    \end{aligned}
\end{equation}
and from which we can extract the first few orders
of the star-exponential 
\begin{align}
    {\cal E}_2(\tau)\rvert_{y=0} & = \tau(\mathscr{E}+\tfrac{R}4)
    +\tfrac{\tau^2}{6}\,R^{ab}\,p_ap_b\,, \\
    {\cal E}_4(\tau)\rvert_{y=0} 
    & = \tfrac{\tau^2}{2}\big(\tfrac{1}{16}
    [R_{abcd} R^{abcd}+R^2] 
    + \mathscr{E}[\mathscr{E} + \tfrac{R}2]
     + \tfrac14 \Omega_{ab}\Omega^{ab}\big) \\ 
    \nonumber & \quad + \tfrac{\tau^3}{3}
    \big(\tfrac{29}{120} R^a{}_{cde} R^{bcde}
    + \tfrac{1}{12} R_{cd} R^{acbd} 
    + \tfrac{1}{20} R^a{}_c R^{bc}
    + \tfrac12 R^{ab}[\mathscr{E}+\tfrac{R}4]\\ 
    \nonumber & \qquad\qquad
    + \tfrac14 \Omega^a{}_c\Omega^{bc}
    + \tfrac12 \nabla_a\nabla_b[\mathscr{E}+\tfrac{R}4]
    + \tfrac{3}{80}\nabla^2 R^{ab}
    + \tfrac{3}{40} \nabla_c\nabla_d R^{acbd}\big)\,p_a p_b \\
    \nonumber & \quad
    + \tfrac{\tau^4}{4}\big(\tfrac{1}{18} R^{ab} R^{cd}
    + \tfrac{7}{45} R^a{}_\times{}^b{}_\times
    R^{c \times d \times} 
    + \tfrac{1}{30} \nabla^a \nabla^b R^{cd}\big)\,
    p_a p_b p_c p_d\,,
\end{align}
using the recursion \eqref{eq:recursion_exp}.
The Heat Kernel coefficients $\mathtt{A}_k[\widehat{H}]$
for $k=0,2,4$ are then simply obtained by integrating
${\cal E}_k(-1)$ respectively on $p$ against
a Gaussian measure as explained previously, leading to
\begin{align}
    \mathtt{A}_0 [\widehat{H}] & = \tfrac{1}{(4\pi)^{d/2}}\,
    \int_\Base {\rm d}^dx\,\sqrt{g}\,\tr_E(\1_E)\,, \\
    \mathtt{A}_2[\widehat{H}] & = \tfrac{1}{(4\pi)^{d/2}}\,
    \int_\Base {\rm d}^dx\,\sqrt{g}\,
        \tr_E\big(\mathscr{E}+\tfrac{R}6\,\1_E\big)\,,\\
    \mathtt{A}_4[\widehat{H}] & = \tfrac{1}{(4\pi)^{d/2}}\,
    \int_\Base {\rm d}^dx\,\sqrt{g}\,
    \tr_E\Big(\tfrac{1}{180} [R^{abcd}R_{abcd}-R^{ab}R_{ab}] 
    + \tfrac12 [\mathscr{E} + \tfrac16 R]^2 \\ \nonumber
    & \hspace{200pt} + \tfrac{1}{12} \Omega_{ab}\Omega^{ab}
    + \tfrac16 \nabla^2[\mathscr{E}+\tfrac15 R]\Big)\,,
\end{align}
which reproduces the standard results for the first 
few coefficients of the Heat Kernel of Laplace-type operator,
see e.g. \cite{Gilkey:1975iq}.

\paragraph{Non-minimal operators.}
A more general class of second order differential operators
than the Laplace-type operators discussed above 
is often considered, class referred to as `non-minimal' operators.
They consist of differential operators acting on sections 
of a vector bundle $E \twoheadrightarrow \Base$ 
whose leading piece is \emph{not} proportional to the identity
on $E$, i.e. a typical such operator takes the form
\begin{equation}\label{eq:non-min}
    \widehat{H} = \mathscr{G}^{ab}\nabla_a \nabla_b 
    + \mathscr{V}^a\nabla_a + \mathscr{E}\,,
\end{equation}
with $\mathscr{G}^{ab}=\mathscr{G}^{ba}$ 
and $\mathscr{G}^{ab} \neq g^{ab} \1_E$,
and all tensors are valued in $\End(E)$.

Let us start by computing the symbol of such an operator:
to do so, we simply compute the quantization of a general
symbol which is at most quadratic in momenta,
\begin{equation}
    f(x,p) = f_0^{ab}(x)\,p_a p_b + \hbar\,f_1^a(x)\,p_a
    + \hbar^2\,f_2(x)\,,
\end{equation}
and find, using the lifts \eqref{eq:symbol_lift}
and \eqref{eq:lift_phi}, that it takes the form
\begin{equation}
    \tfrac1{\hbar^2}\,\widehat{f}\phi
    = \big(f_0^{ab}\,\nabla_a\nabla_b
    + [\nabla_b f_0^{ab} - f_1^a]\,\nabla_a
    + \tfrac14[\nabla_a \nabla_b - R_{ab}]f_0^{ab}
    - \tfrac12\nabla_a f_1^a + f_2\big)\phi\,,
\end{equation}
from which we deduce
\begin{equation}
    f_0^{ab} = \mathscr{G}^{ab}\,,
    \qquad 
    f_1^a = \nabla_b\mathscr{G}^{ab}-\mathscr{V}^a\,,
    \qquad
    f_2 = \mathscr{E} + \tfrac12\nabla_a \mathscr{V}^a 
    + \tfrac14[R_{ab}-3\nabla_a \nabla_b]\mathscr{G}^{ab}\,.
\end{equation}
The star-exponential is computed as previously described,
although one should be careful in the order of the factors
since the symbol is now `matrix-valued'. In our case,
\begin{equation}
    {\cal E}_0(\tau)=1\,,
    \qquad
    {\cal E}_1(\tau)\rvert_{y=0}=\tau\,f_1^a\,p_a\,,
\end{equation}
and
\begin{equation}
    \begin{aligned}
    {\cal E}_2(\tau)\rvert_{y=0} & = \tau f_2
    + \tfrac{\tau^2}{2}\big(f_1^a f_1^b 
    + \tfrac14\{\nabla_c\nabla_d f_0^{ab} 
    + \tfrac23 R_{\times c}{}^a{}_d f_0^{b\times}, f_0^{cd}\}
    -\nabla_c f_0^{da} \nabla_d f_0^{cb}\big)p_a p_b \\
    & \quad + \tfrac{\tau^3}{6}
    \big([\nabla_\times\nabla_\times f_0^{ab} 
    + \tfrac23 R_{e\times}{}^a{}_\times f_0^{be}]
    f_0^{c\times} f_0^{c\times} \\ & \qquad\qquad 
    - \nabla_\times f_0^{ea}\{f_0^{b\times},\nabla_e f_0^{cd}\}
    + \tfrac12 f_0^{\times\times} \nabla_\times f_0^{ab}
    \nabla_\times f_0^{cd}\big)p_a p_b p_c p_d\,,
    \end{aligned}
\end{equation}
where the appearance of anticommutator stems 
from the aforementioned small subtlety that every piece
of the symbol is valued in $\End(E)$, and hence 
the product of the various factor in the previous formula
is the \emph{non-commutative} composition of endomorphisms
of $E$ (that we leave implicit so as to avoid obscuring
the notation by introducing a symbol for this composition).

The very first Heat Kernel coefficient already reveals
the new difficulty one is faced with when dealing with
non-minimal operator: this coefficient is given by
\begin{equation}
    \mathtt{a}_0[\widehat{H}\,] = \int \tfrac{\ddp}{(2\pi)^d}\,
    \tr_E\big(e^{-\mathscr{G}^{ab}p_ap_b}\big)\,,
\end{equation}
and due to the non-commuting nature of $\mathscr{G}^{ab}$, 
we cannot immediately compute the integral over $p$
as a Gaussian integral. Assuming that $\mathscr{G}^{ab}$
takes the form
\begin{equation}
    \mathscr{G}^{ab} = g^{ab}\1_E + \zeta\Pi^{ab}\,,
    \qquad 
    \zeta \geq 0\,,
\end{equation}
with $\Pi^{ab}$ an $\End(E)$-valued symmetric tensor,
and $g^{ab}$ is the metric for which $\nabla$ 
is the Levi--Civita connection (plus the part acting on $E$),
we can factor out a Gaussian exponential, so that in general
we can formally compute the momentum integral and end up with
\begin{equation}
    \mathtt{a}_0[\widehat{H}\,] = \tfrac{1}{(4\pi)^{d/2}}\,
    e^{\frac14\Box_p}\,
    \tr_E\big(e^{-\zeta\Pi^{ab}p_ap_b}\big)\big|_{p=0}\,,
\end{equation}
which is nothing but an operatorial representation
of the Wick contractions (on $p$) of the exponential
of $\Pi^{ab} p_a p_b$. If for instance
the endomorphism-valued symmetric tensor $\Pi^{ab}$
is a projector, in the sense that it obeys
\begin{equation}\label{eq:proj}
    \Pi^{(ab} \Pi^{cd)} = g^{(ab}\,\Pi^{cd)}\,,
\end{equation}
the exponential under the trace simplifies to
\begin{equation}
    e^{-\zeta\Pi^{ab}p_ap_b}
    = \1_E + \tfrac{e^{-\zeta p^2}-1}{p^2}\,\Pi^{ab}p_ap_b\,,
\end{equation}
and the Gaussian integral yields
\begin{equation}
    \mathtt{a}_0[\widehat{H}] = \tfrac{1}{(4\pi)^{d/2}}\,
    \Big(\tr_E(\1_E) - [1-(1+\zeta)^{-d/2}]
    \tfrac{1}{d} \tr_E(\Pi)\Big)\,,
    \qquad 
    \Pi \equiv \Pi^{ab}g_{ab}\,,
\end{equation}
in accordance with \cite{Moss:2013cba}.
Moving on to the next non-vanishing Heat Kernel coefficient,
one would need to use the expression for ${\cal E}_2$ above
and again evaluate a Gaussian, so that 
\begin{equation}
    \mathtt{a}_2[\widehat{H}\,] = \tfrac{1}{(4\pi)^{d/2}}
    e^{\frac14\Box_p}\,\tr_E\Big(e^{-\mathscr{G}^{ab} p_a p_b}
    {\cal E}_2(-1)\rvert_{y=0}\Big)\big|_{p=0}\,.
\end{equation}
Assuming that the operator \eqref{eq:non-min} takes a simpler form,
\begin{equation}
    \widehat{H} = \mathscr{G}^{ab}\,\nabla_a\nabla_b
    + \mathscr{E}\,,
    \InEq{i.e.}
    \nabla_c\mathscr{G}^{bc} = 0 = \mathscr{V}^a\,,
\end{equation}
the expression for ${\cal E}_2$ simplifies to
\begin{equation}
    {\cal E}_2(\tau) = \tau\big(\mathscr{E}
    + \tfrac{1}{4} R_{ab} \mathscr{G}^{ab}\big)
    + \tfrac{\tau^2}{12} R_{ec}{}^a{}_d 
    \{\mathscr{G}^{eb}, \mathscr{G}^{cd}\} p_a p_b
    + \tfrac{\tau^3}{36} R_{ij}{}^a{}_k 
    [\mathscr{G}^{ib}, \mathscr{G}^{jc}] \mathscr{G}^{kd}
    p_a p_b p_c p_d\,.
\end{equation}
Suppose further that
\begin{equation}
    \mathscr{G}^{ab} = \mathscr{G}\,g^{ab}\1_E\,,
\end{equation}
where $\mathscr{G}$ is a non-zero, constant, matrix,
then the second Heat Kernel coefficient reads
\begin{equation}
    \mathtt{a}_2[\widehat{H}] = \tfrac{1}{(4\pi)^{d/2}}
    \tr_E\Big(\mathscr{G}^{-d/2}[\mathscr{E}
    + \tfrac16\,R\,\mathscr{G}]\Big)\,,
\end{equation}
in accordance with, e.g. \cite{Iochum:2016ynh, Iochum:2019gej}.

It seems that our method is capable of handling the case
of non-minimal operators, and it would be interesting
to push it for more general cases, for instance considered
in \cite{Gusynin:1997dc, Avramidi:2000isc}, and derive new results.
One advantage here is that no substantial modification
to the method has to be implemented with respect to the case
of minimal operator, all the difficulty boils down to finding
an efficient way of computing the integral over $p$.
We leave this for future work.

\subsection{Higher-order operators}
\label{sec:higherorder}
Let us also consider a class of higher-order operators with symbol
\begin{align}
    f&= (p^2)^\ell + \dots\,,
\end{align}
where the dots denote lower order terms in $p$.
A very simple comment is that \eqref{higherorder} immediately gives the $\mathtt{a}_0$ coefficient, which was obtained in \cite{Gilkey80,Barvinsky:2021ijq} (up to some overall normalization). Moreover, there is a well-known powerful relation \cite{Fegan:1985,Barvinsky:2021ijq} between the Heat Kernel coefficients of operator $\widehat{f}$ and $(\widehat{f})^\ell$:
\begin{align}
    \mathtt{A}_{2m}[(\widehat{f})^\ell] 
    &= \frac{\Gamma(\tfrac{d-2m}{2\ell})}{\ell\,\Gamma(\tfrac{d-2m}{2})} 
    \mathtt{A}_{2m}[\widehat{f}\,]\,.
\end{align}
This relation can easily be proven once we notice that the lift of symbol of $(\widehat{f})^\ell$ is just $F^{\ast \ell}$ if the lift of the symbol $f$ of $\widehat{f}$ is $F$. Indeed, one can imagine a zeta-function $\zeta_F(s)=\Tr\, F^{\ast(-s)}=\sum_n d_n/\lambda_n^{s}$, where $\lambda_n$ are eigenvalues and $d_n$ are their multiplicities. 
The zeta-function is related to the Heat Kernel in a simple way
\begin{align}
    K_{F}(t)&=\Tr\, e^{-t F}_\ast=\sum_n d_n e^{-t\lambda_n}= \frac{1}{2\pi i} \int \mathrm{d}s\, \Gamma(s) t^{-s} \zeta_{F}(s) \,.
\end{align}
The crucial idea is that $(F^{\ast\ell})^{\ast(-s)}=F^{\ast(-\ell s)}$, i.e. $\zeta_{F^{\ast\ell}}(s)=\zeta_{F}(\ell s)$. Therefore, after some simple change of variables, we find
\begin{align}
    K_{F^{\ast\ell}}(t)&=\Tr\, e^{-t F^{\ast\ell}}_\ast= \frac{1}{2\pi i} \int \mathrm{d}s\, \frac{1}{\ell}\Gamma(s/\ell) t^{-s/\ell} \zeta_{F}(s) \,.
\end{align}
Suppose there is contribution $\mathtt{A}_{2m} t^{-(d/2-m)}$ to the Heat Kernel of $F$. This implies that the zeta-function has a pole 
\begin{align}
    \zeta_F(s)\sim \frac{\mathtt{A}_{2m}}{s-(\tfrac{d-2m}{2})} \frac{1}{\Gamma(\tfrac{d-2m}{2})}\,.
\end{align}
Plugging the same pole into the Heat Kernel expansion of $K_{F^{\ast\ell}}(t)$ we have a contribution of the desired form
\begin{align}
    \frac{\Gamma(\tfrac{d-2m}{2\ell})}{\ell\,\Gamma(\tfrac{d-2m}{2})}\mathtt{A}_{2m}[\widehat{f}]\,t^{-(d-2m)/2\ell}\,.
\end{align}
The proof can be considered standard and it applies since the product of operators is replaced in the Fedosov picture by the star-product of lifts of their symbols.

\subsection{4d}
\label{sec:4d}
Let us compute the Heat Kernel expansion of the conformal Laplace operator, which is the simplest possible operator to feed to our gadget. As it was discussed, its symbol reads
\begin{equation}
    f = p^2 + \tfrac{\hbar^2}{4(d-1)}R\,.
\end{equation}
The relevant terms of its lift $F$ are
\begin{align}
\begin{aligned}
    F & = \big(\eta^{ab} + \tfrac13\,y^c y^d\,
    R_c{}^a{}_d{}^b + \tfrac16\,y^c y^d y^e\,
    \nabla_c R_d{}^a{}_e{}^b \\ & \hspace{50pt}
    + \tfrac1{20}\,y^c y^d y^e y^f\,
    \big[\nabla_c \nabla_d R_e{}^a{}_f{}^b
    + \tfrac{4}{3}\,R^a{}_{cd\times} R^b{}_{ef}{}^\times\big]
    + \dots\big)\,p_a p_b \\
    & \qquad + \tfrac{\hbar^2}{4(d-1)}\,\big(R + y^a\nabla_a R
    + \tfrac12\,y^a y^b\,\big[\nabla_a \nabla_b R 
    + \tfrac13\,(d-1)\,R_a{}^{cde} R_{bcde}\big] + \dots\big)\,,
\end{aligned}
\end{align}
up to fourth order. This symbol has two peculiarities:
first, its leading part $F_0$ is \emph{quadratic} in $p$,
so that $F_2$ only depends on $y$; and second, $F_0$ has no
linear term in $y$. Consequently, when computing 
the star-exponential of $F=F_0+\hbar^2 F_2$ at $y=0$,
one can discard any term with a single $y$-derivative
on $F_0$ (and any number of $p$-derivatives) and any term
with $p$-derivative on $F_2$. 

Now, we need to plug in the expansion of $F$ into the expansion coefficients $\mathcal{E}_{k}$ of the star-product exponent. For $k=0$ there is nothing to compute since $\mathtt{a}_0=\mathcal{E}_0=1$. For $k=1$, one finds
\begin{equation}
    {\cal E}_2\rvert_{y=0} = \tfrac{\tau R}{4(d-1)}
    + \tfrac{\tau^2}{6}\,R^{ab}\,p_a p_b\,.
\end{equation}
On Wick contracting the indices we get $\mathtt{a}_2=0$, which is a well-known result. Now let us have a look at the case $k=2$, and for simplicity,
evaluate ${\cal E}_4$ at $y=0$ directly,
\begin{align}
\begin{aligned}
    {\cal E}_4\rvert_{y=0}
    & = \tfrac{\tau^2}{2}\big(\tfrac1{8(d-1)} \nabla^2 R 
    + \tfrac{R^2}{16(d-1)^2} 
    + \tfrac1{16}\,R^{abcd} R_{abcd}\big) \\
    & \quad + \tfrac{\tau^3}{3}\big(
    \tfrac1{8(d-1)}\,\nabla^a \nabla^b R
    + \tfrac{3}{80}\,\nabla^2 R^{ab} 
    + \tfrac{3}{40}\,\nabla_c \nabla_d R^{acbd} \\
    & \hspace{30pt} + \tfrac{R}{8(d-1)}R^{ab}
    + \tfrac{29}{120}\,R^a{}_{cde} R^{bcde} 
    + \tfrac1{12}\,R^{acbd} R_{cd} 
    + \tfrac{1}{20}\,R^a{}_c R^{bc}\big)\,p_a p_b \\ 
    & \quad + \tfrac{\tau^4}{4}
    \big(\tfrac1{30}\,\nabla^a \nabla^b R^{cd}
    + \tfrac{1}{18}\,R^{ab} R^{cd} 
    + \tfrac{7}{45}\,R^a{}_\times{}^b{}_\times
        R^{c \times d \times}\big)\,p_a p_b p_c p_d\,.
\end{aligned}
\end{align}
Finally, integrating ${\cal E}_4$ over $p$ for $\tau=-t$ leads to $\mathtt{a}_4$ being
\begin{align}
    \int \frac{{\rm d}^dp}{\pi^{d/2}\,t^2}\
    e^{-tp^2}\,{\cal E}_4(-t)\rvert_{y=0}
    & = \tfrac{(d-4)^2}{288(d-1)^2}\,R^2
        + \tfrac1{180}\,R^{abcd} R_{abcd}
        - \tfrac{1}{180}\,R^{ab} R_{ab}
        - \tfrac{d-6}{120(d-1)}\nabla^2 R\,,
\end{align}
which agree with the standard result
(e.g. \cite{Gilkey:1975iq, Parker:1987}). Let us note that the entire $d$-dependence comes from the coefficient $\hbar^2/(4(d-1))$ of the $R$ part. Suppose we begin with the symbol
\begin{equation}
    f = p^2 + \hbar^2 \alpha R\,,
\end{equation}
for some constant $\alpha$. Note that $\alpha=1/12$ for the $4d$ conformal Laplacian. Then, $\mathtt{a}_0=1$, but 
\begin{align}
    \mathtt{a}_2=\frac{(1-12\alpha)}{12}R=\frac{d-4}{288 (d-1)} R\,,
\end{align}
where the last expression is for the conformal Laplacian. Likewise,  
\begin{align}
    \mathtt{a}_4 &= \tfrac{(1-12\alpha)^2}{288}\,R^2
        + \tfrac1{180}\,R^{abcd} R_{abcd}
        - \tfrac{1}{180}\,R^{ab} R_{ab}
        + \tfrac{(20\alpha-1)}{120}\nabla^2 R\,.
\end{align}
Note that the coefficient of the scheme-dependent term $\square R$ turns out to be quite often $1/180$ for the conformally coupled scalar and in our approach as well. There are two anomaly coefficients: the type-B is the coefficient of
\begin{align}
   c&: && C_{\mu\nu,\kappa\lambda} C^{\mu\nu,\kappa\lambda}
    = R_{\mu\nu,\kappa\lambda} R^{\mu\nu,\kappa\lambda}
    - 2\,R_{\mu\nu} R^{\mu\nu} + \tfrac13\,R^2\,,
\end{align}
and the Type-A is the coefficient of the Euler density $E$
\begin{align}
    a&: &&E\equiv \epsilon_{abcd}\, R^{a,b} \wedge R^{c,d}
    = d^4x\,\sqrt{g}\,\big(-R_{\mu\nu,\kappa\lambda}
    R^{\mu\nu,\kappa\lambda} + 4\,R_{\mu\nu} R^{\mu\nu}
    - R^2\big)\,,
\end{align}
With this we get the standard $a=1/360$, $c=1/120$.

\subsection{6d}
\label{sec:6d}
Let us repeat the same exercise in $6d$. There is still one $a$ coefficient of the Type-A anomaly aka the Euler density $E$. It is known that there are $3$ different
type-B Weyl anomalies, which at the same time lead
to three Weyl invariants. The latter read
\cite{Bonora:1985cq, Karakhanian:1994yd}
\begin{equation}
    I_1 = C_{\mu\alpha,\nu}{}^\beta\,
    C^{\mu\sigma,\nu}{}^\lambda\,
    C_{\beta\lambda,\sigma}{}^\alpha\,,
    \qquad
    I_2 = C_{\alpha\beta,}{}^{\mu\nu}\,
    C_{\mu\nu,}{}^{\sigma\lambda}\,
    C_{\sigma\lambda,}{}^{\alpha\beta}\,,
\end{equation}
and
\begin{equation}
    I_3 = C_{\mu\nu,\rho}{}^\lambda\,
    \Big(g_{\lambda\sigma}\,\nabla^2
    + 4\,R_{\lambda\sigma}
    -\tfrac65\,g_{\lambda\sigma}R\Big)\,
    C^{\mu\nu,\rho\sigma}\,,
\end{equation}
where we omitted total derivative terms in the last
expression. Therefore, the complete anomaly can be written as
\begin{align}
    a E+ c_1 I_1+ c_2 I_2+ c_3 I_3 +\text{total derivative}\,.
\end{align}
The anomaly for the conformally coupled scalar in $6d$
has been computed in, e.g. \cite{Bastianelli:2000dw, Bastianelli:2000rs, Bastianelli:2000hi},
and for a Dirac fermion in e.g. \cite{Bastianelli:2001tb}.

The first Heat Kernel coefficients $\mathtt{a}_0$, $\mathtt{a}_2$, $\mathtt{a}_4$ are exactly the same as computed before, of course. A word of warning is that one has to set $d=6$ in the formulas of the previous section, e.g. $\alpha=1/20$ instead of $\alpha=1/12$. In particular with $\alpha=1/(4(d-1))$ one reproduces the results of \cite{Parker:1987}.

For the anomaly coefficients we get the expected\footnote{With the normalization of the Euler density given by EulerDensity command of xTras.} $a=1/9072$, and $c_i=(28,-5,-6)/15120$. It might also be interesting to record the total derivative term $\nabla_a J^a$ that we obtain
\begin{align}
\begin{aligned}
    J^a&=- \tfrac{1}{840} R^{bc} \nabla^{a}R_{bc} -  \tfrac{1}{25200} R \nabla^{a}R + \tfrac{1}{280} R^{bcdm} \nabla^{a}R_{bdcm} + \\
    &\qquad +\tfrac{1}{4200} \nabla_{b}\nabla^{b}\nabla^{a}R -  \tfrac{1}{12600} R^{a}{}_{b} \nabla^{b}R -  \tfrac{1}{1260} R^{a}{}_{bcd} \nabla^{d}R^{bc}\,.
\end{aligned}
\end{align}

\subsection{8d}
\label{sec:8d}
To illustrate the efficiency of new approach let us compute the conformal anomaly of the conformally-coupled scalar field in $8d$.\footnote{In principle, the $\mathtt{a}_8$ coefficient for a Laplace-type operator is available in some form, see \cite{Amsterdamski:1989bt, Avramidi:1990ug}.} First of all, there are quite a few conformal invariants in $8d$. The local conformal invariants, i.e. strictly Weyl-invariant densities, have been classified in \cite{Boulanger:2004zf}.\footnote{We are extremely grateful to Nicolas Boulanger for sharing his Mathematica notebook with these conformal invariants, which has saved us a lot of time. } The general statement is that global conformal invariants are given by the local ones plus total derivatives (plus the Euler term) \cite{Deser:1993yx,Alexakis2012,Boulanger:2018rxo}. Note that there is a slight difference between conformal invariants and conformal anomalies, see e.g. \cite{Bonora:1983ff,Boulanger:2007ab,Boulanger:2018rxo}. The local conformal invariants can be arranged by the number of derivatives that the Weyl tensors carry. (0) There are $7$ zero-derivative invariants that are scalars built from four Weyl tensors that date back to \cite{Fulling:1992vm} (the expressions are taken from \cite{Boulanger:2025oli}):
\begin{align}
\begin{aligned}
I_6 &=
C_{ab}{}^{ef}\,
C^{abcd}\,
C_{ce}{}^{gh}\,
C_{dfgh}\,,
& 
I_7 &=
C_{a}{}^{e}{}_{c}{}^{f}\,
C^{abcd}\,
C_{b}{}^{g}{}_{d}{}^{h}\,
C_{egfh}\,,
\\[2mm]
I_8 &=
C_{ab}{}^{ef}\,
C^{abcd}\,
C_{cd}{}^{gh}\,
C_{egfh}\,,
&
I_9 &=
C_{abc}{}^{e}\,
C^{abcd}\,
C_{d}{}^{fgh}\,
C_{egfh}\,,
\\[2mm]
I_{10} &=
C_{abcd}\,
C^{abcd}\,
C_{efgh}\,
C^{egfh}\,,
&
I_{11} &=
C_{a}{}^{e}{}_{c}{}^{f}\,
C^{abcd}\,
C_{b}{}^{g}{}_{f}{}^{h}\,
C_{dheg}\,,
\\[2mm]
I_{12} &=
C_{ac}{}^{ef}\,
C^{abcd}\,
C_{b}{}^{g}{}_{e}{}^{h}\,
C_{dhfg}\,.
\end{aligned}
\end{align}
(4) There is a single ``four-derivative'' invariant $I_1$, whose ``leading symbol'' is $C\square^2 C$. (2) There are four ``two-derivative'' invariants $I_{2,3,4,5}$, whose ``leading symbols'' contain $C(\nabla C)(\nabla C)$ and $C C(\nabla^2 C)$. It was found in \cite{Chen:2024kuw} that $I_{4,5}$, being local conformal invariants, globally, i.e. modulo total derivatives, can be expressed in terms of the other invariants. Therefore, as far as global conformal anomalies are of interest, the basis is formed by $I_{1,2,3}$ and $I_{i}$, $i\in[6,12]$. 

In terms of anomalies, the latter means that we have ten Type-B anomaly coefficients, which we denote $c_i$, $i\in \{1,2,3\}\cup[6,12]$ with a ``gap'' in $4,5$ in deference to the highly nontrivial results of \cite{Boulanger:2004zf}. There is also a single Type-A anomaly coefficient $a$. Of course, one could choose some other basis for type-B anomalies, but the one of \cite{Boulanger:2004zf} seems to be more motivated at the moment, since the expressions are the simplest ones in terms of Weyl-covariant derivatives (modulo total derivatives). Any generic conformally-invariant expression in terms of (derivatives of) Riemann tensor would respect Weyl symmetry less manifestly. To return to our discussion, the most general ansatz for an $8d$ anomaly reads
\begin{align}
    \sum_{i\in [1,3]\cup[6,12]} c_i I_i +a E +\text{total derivative}
\end{align}

In principle, nothing changes in the general formula for Heat Kernel \eqref{bestHeatKernel} as we climb up to $8d$, just the number of terms of order $\hbar^8$ grows as compared to $\hbar^6$ in $6d$. There are several features of the Heat Kernel expansion of the conformal Laplace operator in our approach that simplify the calculation. Let us concentrate on the anomaly part. Firstly, $\pl_a\pl^bF$ and $\pl_a\pl^{bc}F$ vanish at $y=0$ and, hence, all terms involving them can be removed from $\mathcal{E}_k$. Secondly, the derivative $\pl_{a_1 \dots a_k}\pl^{b_1 \dots b_n}$ with $n>2$ vanish since $F$ is no more than quadratic in $p$. Thirdly, there are some simple restrictions by degree: assume that each $\nabla$ contributes $1$ and each $R$ contributes $2$, the anomaly is the degree $d$ piece. However, $\pl_a F|_{y=0}\sim \nabla_a R$ has degree $3$ already and the lowest degree of terms in $\pl_{a_1 \dots a_k}\pl^{b_1 \dots b_n}F|_{y=0}$ is $k$ for $k>1$ and $n=0,1,2$. Therefore, there are quite a few terms $\pl \dots F \pl F \dots \pl \dots F$ in $\mathcal{E}_k$ that do not contribute to the anomaly. With this in mind a straightforward calculation leads to 
\begin{align}\notag
    c_i=\left\{\frac{1}{4536},-\frac{19}{226800},-\frac{7}{54000},-\frac{259}{32400},-\frac{179}{113400},-\frac{139}{453600},\frac{127}{18900},\frac{17}{302400},\frac{13}{9450},-\frac{274}{14175}\right\}
\end{align}
and the value of the $a$-anomaly is 
\begin{align}
    a&= \frac{23}{48\cdot 113400}\,.
\end{align}
Maybe a slightly more compact way to write the $c$-anomalies is 
\begin{align}
    c_i=\frac{1}{226800}
    \left\{
    50,\,-19,\,-\frac{147}{5},\,-1813,\,-358,\,-\frac{139}{2},
    \,1524,\,\frac{51}{4},\,312,\,-4384
    \right\}.
\end{align}
There is a nontrivial total derivative term that allows us to expand the Heat Kernel coefficient in the chosen basis of anomalies. 
Let us note that the Weyl anomaly $a$-coefficient is known for the conformally-coupled scalar field in all dimensions \cite{Casini:2010kt} and we agree up to the normalization of the Euler term. Another consistency check is to recall that $c_1$ is associated with the highest derivative term
\begin{align}
    I_1&= q\, C_{ab,cd} \square^2 C^{ab,cd}+\dots
\end{align}
where $q$ is some number. Therefore, it is responsible for the $c_T$ coefficient of the stress-tensor two-point function
\begin{align}
    \langle T_{ab} T_{cd}\rangle&= \frac{c_T}{S^2_d} \frac{I_{ab,cd}}{x^{2d}}\,.
\end{align}
The two-point function is not well-defined as a distribution: $1/x^{16} \sim \square^4 \delta^8(x) $. This is exactly the anomaly that is captured by $I_1$. There are several terms in $I_1$ that lead to $C\square^2 C$ in the flat space approximation.\footnote{One way to find the right coefficient is to use the decomposition of the $Q_8$-curvature in terms of the $I_\bullet$-basis, which was obtained in \cite{Boulanger:2025oli}. It gives $Q_8 \ni -5/3 I_1$. On the other hand, it is well-known that $Q_d\ni -\tfrac{(d-2)}{8(d-3)} C (-\square)^{d/2-2} C$, which implies that the coefficient of $C\square^2 C$ is given by $9/100 c_1$, i.e. $q=9/100$. } The canonical normalization is \cite{Aros:2026gms}
\begin{align}
    (4\pi)^4 \langle T_m{}^{m}\rangle & \ni -a E + c\frac{d(d-2)}{8(d-3)}\, C\square^{d/2-2} C\, \Big|_{d=8}\ni c\frac{6}{5}\, C\square^{2} C 
\end{align}
This gives $c=3/40 c_1 =1/60480$, which is the correct value of the central charge of the free scalar field \cite{Aros:2026gms} (it gives $c_T=8/7$, which is $d/(d-1)$ \cite{Osborn:1993cr}). Likewise, $c_{2,3}$ should manifest themselves in three-point functions $\langle TTT\rangle$. There is not much meaning in individual values of $c_{6,\dots,12}$ since one could choose another basis among the algebraic invariants, but they should contribute to four-point functions $\langle TTTT\rangle$. We see that $a/c=23/90$, which tells us that the free scalar field is not captured by just Einstein--Hilbert action, for which $a/c=1$. In principle, with the help of \cite{Chen:2024kuw} one could determine the structure of the higher derivative corrections to Einstein--Hilbert action for the AdS/CFT dual of the free scalar field. To conclude, with $c_T$ and $a$ correctly reproduced, we have enough confidence in the rest of the anomaly coefficients.\footnote{The only thing that prevents us from going to $10d$ is the lack of the basis of conformal invariants. Let us note that nothing in the practical implementation of our procedure would require an advanced package like xAct. Nevertheless, xAct and xTras are extremely useful in dealing with the final answer, e.g. to decompose it in a given basis of conformal invariants. }

\section{Conclusions and Discussion}
\label{sec:conclusions}
In the paper we proposed a new approach to Heat Kernel. The main advantages include: (1) manifest general covariance at all stages; (2) manifest dimension independence (till Wick's contractions for higher order operators); (3) closed form answers for the coefficients; (4) easy to implement practically; (5) generality, e.g. lower- and higher- order operators are treated in the same way.\footnote{To be fair, some of these features are also present in other approaches, but not all together and not in a way that would give explicit closed form answers.} To illustrate our approach we have computed the first few Heat Kernel coefficients of the conformally-coupled scalar field, which corresponds to conformal anomalies in $2$, $4$, $6$ and $8$ dimensions. The results in eight dimensions are new and pass some nontrivial checks.

Concerning the $d$-independence, it is clear that the Fedosov connection $A$ and the lift $F$ of a generic symbol $f$ do not have the dimension $d$ of the spacetime manifold in them, as well as $\exp_\ast [-tF]$ does not. Indeed, they are constructed by iterating contractions via the Poisson bracket, which does not depend on the metric structure/dimension. Note that a specific symbol, e.g. that of the conformal Laplace, might have $d$ in it.

In principle, the complete covariant Heat Kernel $\exp_\ast[-t H]$ is also available in our approach. So far we have just concentrated on the short time expansion of its trace. The off-diagonal piece of the Heat Kernel $K(t;x,x')$ can be computed by tracing $\exp_\ast[-t H]$ with $|x\rangle\langle x'|$. Therefore, it would be interesting to study another regime, e.g. a small curvature expansion but with $t$ taken into account exactly. 

Some obvious directions for the future include: (1) higher order differential operators: (a) GJMS-operators, including their generalization for fermions \cite{Holland:2001, Fischmann:2013, Fischmann:2014}; (b) conformal operators for tensor fields; (2) other cases with a nontrivial bundle component; (3) it might be interesting to consider supersymmetric realizations of the approach or to see if there exist more efficient formulations in specific lower dimensions $4$, $6$, $8$, $10$. 

It should be easy to generalize the approach to non-commutative field theories. The only change that needs to be done is to add a genuine non-commutativity on the base manifold, which can again be implemented via the Fedosov approach. Another direction is to include boundary effects, e.g. Dirichlet vs. Neumann boundary conditions, which are important in AdS/CFT applications, see e.g. \cite{Barvinsky:2005ms,Hartman:2006dy}.

The Fedosov approach dates back to 1985\footnote{This is the year the original paper was published in Russian.} \cite{Fedosov:1994zz, Fedosov:1996} and it does give a simple constructive formula for the star-product on arbitrary symplectic manifold, e.g. on a cotangent bundle \cite{Fedosov:2001}. Nevertheless, the problem of constructing the trace\footnote{The original Fedosov approach was to appeal to the existence of Darboux coordinates at least locally.} was fully solved much later by Feigin, Felder and Shoikhet \cite{Feigin:2005}, who constructed an appropriate measure $\mu(F|A,\dots,A)$. Interestingly, their result was a consequence of Shoikhet's proof \cite{Shoikhet:2000gw} of Shoikhet--Tsygan--Kontsevich formality. We observe that for the case of cotangent bundles the measure boils down to the most naive $\sqrt{g}$, thereby not resulting in any additional complications in Heat Kernel expansion. Nevertheless, it is still remarkable that our construction would have been incomplete had it not been for (Shoikhet--Tsygan--)Kontsevich formalities.

\section*{Acknowledgment}
We are grateful to Xavier Bekaert, Nicolas Boulanger, Maxim Grigoriev, Euihun Joung, Rodrigo Olea, Alexey Sharapov for lots of closely related useful discussions over the years. We would like to thank the organizers of Geometry for Higher Spin Gravity: Conformal Structures, PDEs, and Q-manifolds, ESI, Vienna, 2021 and Workshop “Conformal higher spins, twistors and boundary calculus”, UMONS, 2025 for the very stimulating atmosphere. T. B. is grateful to Kevin Grosvenor for useful discussion about the covariant Fourier transform and its use in Heat Kernel computations. A lot of the presented ideas originated from our attempts to develop conformal higher-spin theories as introduced by Arkady Segal \cite{Segal:2002gd} and Arkady Tseytlin \cite{Tseytlin:2002gz}, see also the important contribution \cite{Bekaert:2010ky} that binds together these two proposals and relies on Heat Kernel expansion. For manipulations with various expressions of conformal anomalies we used xAct \cite{xAct,Martin-Garcia:2008ysv} and xTras \cite{Nutma:2013zea}. In particular, without reinventing the wheel, the normalization of the Euler density is given by the EulerDensity function of xTras. The work of T. B. and E. S. was partially supported by the European Research Council (ERC) under the European Union’s Horizon 2020 research and innovation programme (grant agreement No 101002551). E.S. is a research associate of the Fonds de la Recherche Scientifique – FNRS. 

\appendix

\providecommand{\href}[2]{#2}\begingroup\raggedright\endgroup

\end{document}